\let\ifarxiv=\iftrue     
\pdfoutput=1
\documentclass[12pt,a4paper]{article}

\ifarxiv\ifnum\pdfoutput=1\else
\PassOptionsToPackage{hypertex}{hyperref}
\PassOptionsToPackage{draft}{graphicx}
\usepackage{showkeys}
\fi\fi

\usepackage{amsmath,amssymb}
\usepackage[bookmarks=true,hyperfigures=true]{hyperref}
\usepackage{graphicx}
\usepackage[nosort]{cite}
\usepackage[bulletsep]{collref}
\usepackage{feynmp}
\usepackage{multirow}
\usepackage{listings}
\usepackage{longtable}
\usepackage{multicol}
\usepackage{hhline}

\usepackage[a4paper,text={173mm,216mm},centering]{geometry}

\allowdisplaybreaks[3]

\usepackage[font=small,labelfont=bf,width=0.85\textwidth]{caption}

\makeatletter
\let\old@startsection=\@startsection
\renewcommand{\@startsection}[6]{\old@startsection{#1}{#2}{#3}{#4}{#5}{#6\mathversion{bold}}}
\makeatother

\makeatletter
\newlength{\apb@width}
\newcommand{\autoparbox}[2][c]{\settowidth{\apb@width}{#2}\parbox[#1]{\apb@width}{#2}}

\makeatother

\let\oldPhi=\Phi
\let\oldPsi=\Psi
\let\oldGamma=\Gamma
\let\oldDelta=\Delta
\let\oldSigma=\Sigma
\let\oldLambda=\Lambda
\let\oldTheta=\Theta
\let\oldPi=\Pi
\let\oldXi=\Xi
\let\oldUpsilon=\Upsilon
\let\oldOmega=\Omega
\renewcommand{\Phi}{\mathnormal{\oldPhi}}
\renewcommand{\Psi}{\mathnormal{\oldPsi}}
\renewcommand{\Gamma}{\mathnormal{\oldGamma}}
\renewcommand{\Sigma}{\mathnormal{\oldSigma}}
\renewcommand{\Delta}{\mathnormal{\oldDelta}}
\renewcommand{\Theta}{\mathnormal{\oldTheta}}
\renewcommand{\Lambda}{\mathnormal{\oldLambda}}
\renewcommand{\Pi}{\mathnormal{\oldPi}}
\renewcommand{\Xi}{\mathnormal{\oldXi}}
\renewcommand{\Upsilon}{\mathnormal{\oldUpsilon}}
\renewcommand{\Omega}{\mathnormal{\oldOmega}}

\ifx\genfrac\sdflkaj\else\fi
\newcommand{\sfrac}[2]{{\textstyle\frac{#1}{#2}}}
\newcommand{\half}{\sfrac{1}{2}}

\newcommand{\alg}[1]{\mathfrak{#1}}

\newcommand{\abs}[1]{|#1|}

\newcommand{\be}{\begin{equation}}
\newcommand{\ee}{\end{equation}}

\newcommand{\Tr}{\mathop{\mathrm{Tr}}}

\newcommand{\nn}{\nonumber}

\def\[{\begin{equation}}
\def\]{\end{equation}}
\def\<{\begin{eqnarray}}
\def\>{\end{eqnarray}}

\makeatletter
\def\mr@ignsp#1 {\ifx\:#1\@empty\else #1\expandafter\mr@ignsp\fi}%
\newcommand{\multiref}[1]{\begingroup
\xdef\mr@no@sparg{\expandafter\mr@ignsp#1 \: }%
\def\mr@comma{}%
\@for\mr@refs:=\mr@no@sparg\do{\mr@comma\def\mr@comma{,}\ref{\mr@refs}}%
\endgroup}
\makeatother

\newcommand{\eqn}[1]{(\ref{#1})}

\ifx\href\asklfhas\newcommand{\href}[2]{#2}\fi

\newcommand{\hypref}[2]{\ifx\href\asklfhas #2\else\href{#1}{#2}\fi}

\renewcommand{\eqref}[1]{(\multiref{#1})}

\newcommand{\hl}{\hhline{*{3}{|-}||*{3}{-|}||-|}}
\newcommand{\hld}{\hhline{*{3}{:=}::*{3}{=:}::=:}}

\newcommand{\e}{\ensuremath{\varepsilon}}

\newcommand{\soll}{\stackrel{!}{=}}

\newcommand{\tr}[1]{\mathrm{Tr}\left(#1\right)}

\newcommand{\bleq}{\ensuremath{\mathrel{\phantom{=}}}}

\newcommand{\kla}[1]{\left( #1 \right)}

\newcommand{\ekla}[1]{\left[ #1 \right]}
\newcommand{\skla}[1]{\left\langle #1 \right\rangle}

\newcommand{\dv}[1]{\ensuremath{\text{d}^4#1} \,}

\newcommand{\indexfett}[2][f]{\if#1f{\index{#2|textbf}}\else{\index{#2}}\fi}

\newcommand{\lnk}[1]{\ln \frac{\e^2}{x_{#1}^2}}
\newcommand{\lnl}[3]{\ln \frac{\e^2 x_{#1}^2}{x_{#2}^2 x_{#3}^2}}

\newcommand{\Op}{\mathcal{O}}
\newcommand{\cO}{\mathcal{O}}
\newcommand{\K}{\mathcal{K}}
\newcommand{\D}{\mathcal{D}}

\newcommand{\Otil}{\widetilde{\mathcal{O}}}

\newcommand{\Dn}{\Delta^{(0)}}
\newcommand{\Df}{\Delta}

\newcommand{\nnl}{\nonumber\\}

\newcommand{\ps}{$\hspace{-6pt}\phantom{\Big]}$}
\newcommand{\oc}[1]{\raisebox{-1pt}{#1}}
\newcommand{\psx}{\\[4pt]}
\newcommand{\ocx}[1]{\raisebox{-3pt}{#1}}

\renewcommand{\fmfdot}[1]{\fmfv{decor.shape=circle,decor.filled=full,
decor.size=1.5thick}{#1}}

\ifx\hypersetup\sadfkjashdfkxja\newcommand\hypersetup[1]{}\fi

\hypersetup{plainpages=false}
\hypersetup{pdfpagemode=UseNone}
\hypersetup{bookmarksnumbered=true}
\hypersetup{pdfstartview=FitH}
\hypersetup{colorlinks=false}
\hypersetup{citebordercolor={.5 1 .5}}
\hypersetup{urlbordercolor={.5 1 1}}
\hypersetup{linkbordercolor={1 .7 .7}}

\newcommand\oplength{\Delta^{(0)}}

\begin{document}
\thispagestyle{empty}

\ifarxiv\else\begingroup\raggedleft\footnotesize\ttfamily
HU-EP-10/nn\\
\endgroup\fi

\begingroup\centering
{\Large\bfseries\mathversion{bold} One-Loop Spectroscopy of Scalar
Three-Point Functions in planar $\mathcal{N}=4$ super Yang-Mills Theory\par}%
\hypersetup{pdfsubject={}}%
\hypersetup{pdfkeywords={}}%
\ifarxiv\vspace{15mm}\else\vspace{15mm}\fi

\hypersetup{pdfauthor={}}%
\begingroup\scshape\large 
 Andr\'e Gro{\ss}ardt
and Jan Plef\/ka
\endgroup
\vspace{5mm}

\begingroup\ifarxiv\small\fi
\textit{Institut f\"ur Physik, Humboldt-Universit\"at zu Berlin, \\
Newtonstra{\ss}e 15, D-12489 Berlin, Germany}\\[0.2cm]
\ifarxiv\texttt{\{andre.grossardt,jan.plefka\}@physik.hu-berlin.de\phantom{\ldots}}\fi
\endgroup
\vspace{4cm}

\textbf{Abstract}\vspace{5mm}\par
\begin{minipage}{14.7cm}
We report on a systematic study of scalar field three-point functions in 
planar $SU(N)$ $\mathcal{N}=4$ super Yang-Mills theory at the one-loop level. 
For this we have computed
a sample of 70 structure constants  at one-loop order involving
primary operators  of up to and including length five built entirely from scalar fields. We observe in 
all 17 cases occurring in our sample that 
the one-loop structure constant of two protected
chiral primary operators and one unprotected operator is given by a simple linear function involving the
anomalous scaling dimension of the latter. Moreover, a 
similar simple one-loop formula is proven for 
the three-point structure constants of the Konishi operator and two arbitrary protected or
un-protected  operators. It is again determined by the anomalous scaling dimensions of the
operators involved.
\end{minipage}\par
\endgroup 
\newpage


\setcounter{tocdepth}{2}
\hrule height 0.75pt
\tableofcontents
\vspace{0.8cm}
\hrule height 0.75pt
\vspace{1cm}

\setcounter{tocdepth}{2}

\section{Introduction and Conclusions}

Following the discovery of integrable structures \cite{Minahan:2002ve, Beisert:2003tq, Bena:2003wd,Beisert:2003yb} in the AdS/CFT correspondence
\cite{Maldacena:1998re,Witten:1998qj,Gubser:1998bc}
our understanding of 
${\mathcal N}=4$ supersymmetric Yang-Mills (SYM) theory \cite{Brink:1976bc, Gliozzi:1976qd} 
and the dual $AdS_5\times S^5$ superstring theory 
has greatly advanced. To a large extent this progress
occurred in the problem of finding the exact all-loop form of the 
anomalous scaling dimensions of local gauge
invariant operators of the gauge theory alias 
the spectrum of string excitations in the string model.
The key was a mapping of the problem to an integrable spin chain which emerged from a one-loop
perturbative study of the diagramatics involved by Minahan and Zarembo
\cite{Minahan:2002ve}. Moving on to higher loops
the spectral problem was mapped to the diagonalization of a long-range spin chain model, 
whose precise microscopic form remains unknown 
\cite{Beisert:2003yb,Beisert:2003ys}. Nethertheless 
assuming integrability the spin-chain S-matrix
could be algebraically constructed and the spectral problem was rephrased for asymptotically long 
operators to the solution of a set of nested Bethe equation 
\cite{Staudacher:2004tk,Beisert:2005tm} (for reviews see 
\cite{Tseytlin:2004cj,Belitsky:2004cz,Zarembo:2004hp,Plefka:2005bk,Minahan:2006sk,Arutyunov:2009ga,Beisert:2004yq,Beisert:2004ry}). 
The central remaining problem is now the understanding of wrapping interactions, 
which affect short operators at lower loop orders \cite{Ambjorn:2005wa}, 
\cite{Bajnok:2009vm,Fiamberti:2009jw}. From the algebraic
viewpoint
important progress was made by thermodynamic Bethe ansatz techniques 
\cite{Gromov:2009bc,Bombardelli:2009ns,Arutyunov:2009ur}
which
also lead to a conjecture for the exact numerical scaling dimensions of the 
Konishi operator, the shortest unprotected operator in the theory \cite{Gromov:2009zb,Frolov:2010wt}.

Next to the scaling dimensions there also exist remarkable all-order results in planar 
${\mathcal N}=4$ SYM for supersymmetric
Wilson-loops of special geometries \cite{Erickson:2000af,Drukker:2000rr}
as well as for scattering amplitudes of four and five external
particles \cite{Bern:2005iz,Drummond:2008vq}, being closely related to light-like Wilson lines
\cite{Alday:2007hr}, see \cite{Alday:2008yw,Henn:2009bd} for reviews.

Given these advances in finding exact results it is natural
to ask if one can make similar statements for three-point functions of local gauge invariant
operators. Due to conformal symmetry the new data appearing are the structure constants 
which have a nontrivial coupling constant $\lambda=g^{2}N$ dependence and also appear in the
associated operator product expansion. In detail we have for renormalized operators
\be
 \skla{\Otil_\alpha(x_1) \, \Otil_\beta(x_2) \, \Otil_\gamma(x_3)} = \frac{C_{\alpha \beta \gamma}}{\abs{x_{12}}^{\Df_\alpha + \Df_\beta - \Df_\gamma} \abs{x_{23}}^{\Df_\beta + \Df_\gamma - \Df_\alpha} \abs{x_{13}}^{\Df_\alpha + \Df_\gamma - \Df_\beta} \abs{\mu}^{\gamma_\alpha + \gamma_\beta + \gamma_\gamma} } \, ,
\label{3ptconvn}
\ee
where $\Delta_{\alpha}=\Dn_{\alpha}+\lambda\,\gamma_{\alpha}$ 
denotes the scaling dimensions of the operators involved
with $\Dn$ the engineering and $\gamma$ the anomalous scaling dimensions, $\mu$ the renormalization
scale and
\be
 C_{\alpha \beta \gamma} = C^{(0)}_{\alpha \beta \gamma} + \lambda \,
 C^{(1)}_{\alpha \beta \gamma} + O(\lambda^2)\, 
\label{cdef}
\ee
is the scheme independent structure constant representing the new observable arising in three-point
functions one would like to find. Similar to the case of two-point functions there are non-renormalization
theorems for three-point correlation functions of chiral primary (or 1/2 BPS)
operators, whose structure constants do not receive radiative corrections
\cite{Eden:1999gh,Arutyunov:2001qw,Heslop:2001gp,Lee:1998bxa,Basu:2004nt}.

The study of three-point functions involving non-protected operators allowing for 
a non-trivial coupling constant dependence of the structure constants is still largely in its infancy. 
Direct computations of three-point functions are 
\cite{Bianchi:2001cm,Beisert:2002bb,Roiban:2004va,Okuyama:2004bd,Alday:2005nd,Alday:2005kq,Georgiou:2009tp} while 
\cite{Arutyunov:2000im} analyzed the problem indirectly through an OPE decompostition
of four-point functions of chiral primaries. The works
\cite{Beisert:2002bb,Georgiou:2009tp} focused on non-extremal correlators involving 
scalar two-impurity operators which are particularly relevant in the BMN limit. 
The mixing problem of these operators with fermion and derivative impurities was analyzed in
\cite{Georgiou:2008vk}.
\cite{Roiban:2004va} considered extremal correlators of a very special class of operators 
allowing an interesting map to spin-chain correlation functions, while 
\cite{Casteill:2007td} addresses similar questions from the perspective of the non-planar
contribution of the dilatation operator.

The two works \cite{Okuyama:2004bd,Alday:2005nd} considered the general problem of finding the structure constants
of scalar field primary operators discussing important aspects of scheme independence for the
determination of $C^{(1)}_{\alpha\beta\gamma}$.
In this paper we shall continue this work and report on a 
systematic one-loop study of short single trace conformal primary operators built from the six
real scalar fields of the theory in the planar limit. For this we developed a combinatorial
dressing technique to promote tree-level non-extremal three-point correlation functions to
the one-loop level which is similar to the results reported in \cite{Alday:2005nd}.
This is then used to compute a total of 70 structure constant at the one-loop
level involving 11 different scalar field conformal primary operators up to and including
length five.
The restriction to this particular set of operators arose from the necessity to lift the
operator degeneracy in the scalar sector by diagonalising the two-point functions at one-loop.
However, the mixing problem in the sector with fermionic and derivative insertions was not
resolved, which in general contributes structure constants at the ${\cal O}(\lambda)$ level.
The main motivation for this spectroscopic study is to provide data to
test and develop future conjectures on the form of the three-point structure constants potentially
making use of integrability.

Next to providing this one-loop data 
two general observations could be made. Firstly, in all cases that we computed
the structure constants  involving two protected ($1/2$ BPS or chiral primary)
operators with an unprotected operator follow a simple linear expression in the anomalous
scaling dimensions. In the normalization conventions of \eqn{3ptconvn} and \eqn{2ptren}
for renormalized  operators we find the relation
\be
\frac{C^{(1)}_{\alpha \beta \gamma,\, \text{non-extremal}}}{C^{(0)}_{\alpha \beta \gamma,\, \text{non-extremal}}}=
-\frac{1}{2}\, \gamma_{\gamma}\, \qquad \text{if } \gamma_{\alpha}=\gamma_{\beta}=0\, ,
\label{eq1}
\ee
in all 17 cases that occurred in our study. 
It should be stressed, however, that possible additional contributions to 
$C^{(1)}_{\alpha \beta \gamma,
\, \text{non-extremal}}$
arise from operator mixing at the ${\cal O}(\sqrt{\lambda})$ order with 
fermionic insertions or covariant derivative insertions at
the ${\cal O}(\lambda)$ level respectively, as was 
studied in \cite{Georgiou:2008vk,Georgiou:2009tp}\footnote{We thank 
the authors of these papers for pointing this out to us.}. These additional contributions due to operator-mixings
beyond the $SO(6)$ sector have not been taken into account here and might change the above result.

Secondly, for the non-extremal three-point correlator of the Konishi operator
${\cal K}=\Tr(\phi^{i}\phi^{i})$ with two arbitrary scalar field primary operators we prove the relation
\be
\frac{C^{(1)}_{\alpha \beta \K,\, \text{non-extremal}}}{ C^{(0)}_{\alpha \beta \K,\, \text{non-extremal}}} = - \delta_{\alpha\beta}\, 
\kla{2\frac{\gamma_\alpha}{\Dn_\alpha}  + \frac{\gamma_\K}{\Dn_\K}} \, .
\ee
Note that this result is in accordance with \eqn{eq1} for 
$\gamma_{\alpha}=0$ as $\Dn_\K=2$. It is important to stress
that both results only apply for non-extremal correlation functions. Extremal correlation functions
are such that $\Dn_{\gamma}=\Dn_{\alpha}+\Dn_{\beta}$ i.e.~the length of the longest operator
is equal to the sum of the two shorter ones. Here there also exists a compact one-loop
formula due to Okuyama and Tseng \cite{Okuyama:2004bd} see equation
\eqn{eqn:form of structure constants for extremal correlators}.

It would be very interesting to see whether these simple structures are stable at higher loop-order 
and also for non-purely scalar field primary operators such as the twist $J$ operators for example. 
Even more interesting would be a computation of three-point functions involving non-protected operators
at strong coupling via classical string theory. Here very interesting first steps were done by constructing
suitable spinning string solution in \cite{Janik:2010gc} approaching the boundary of $AdS_{5}$ 
and in the construction of classical string vertex operators \cite{Buchbinder:2010vw}.

\section{General structure and scheme dependence of two and three-point functions}

We want to compute planar two- and three-point functions of local scalar operators 
at the one-loop order.
For this it is important to identify the regularization scheme independent information.

To begin with a scalar two-point function of bare local operators $\Op_{\alpha}^{B}(x)$ 
in a random basis can be brought into diagonal form under a suitable linear transformation 
$\Op_\alpha = M_{\alpha \beta} \Op^B_\beta$ with a coupling constant 
$\lambda=g^{2}\,N$ independent mixing matrix $M_{\alpha\beta}$ as we are working at the one-loop
level\footnote{Note that the two-loop diagonalization will involve a mixing matrix proportional to $\lambda$.}
\begin{equation}
 \skla{ \Op_\alpha(x_1) \, \Op_\beta(x_2) } = \frac{\delta_{\alpha \beta}}{x_{12}^{2 \Dn_\alpha}} \kla{ 1 + \lambda \, g_\alpha - \lambda\, \gamma_\alpha \,\ln \abs{x_{12} \epsilon^{-1}}^2 }\, ,
\qquad x_{12}^{2}:=(x_{1}-x_{2})^{2}\, ,\end{equation}
where $\epsilon$ represents a space-time UV-cutoff and $\Delta^{(0)}_{\alpha}$ the engineering
scaling dimension of $\cO_{\alpha}$. Clearly the finite contribution
to the one-loop normalization $g_{\alpha}$ is scheme dependent \cite{Okuyama:2004bd,Alday:2005nd}
as a shift in the cutoff parameter 
$\epsilon\to e^{c}\,\epsilon$ changes 
\be
g_{\alpha}\to g_{\alpha}+ 2\, c\, \gamma_\alpha\, .
\ee

One may now define the renormalized operators via
\begin{equation}
 \Otil_\alpha = \Op_\alpha \kla{ 1 - \frac{\lambda}{2} g_\alpha - \lambda \gamma_\alpha \ln \abs{\mu\, \epsilon} + O(\lambda^2) } \label{eqn:Renormalized ops}
\end{equation}
with a renormalization momentum scale $\mu$ to
obtain finite canonical two-point correlation
functions
\begin{equation}
\label{2ptren}
\skla{ \Otil_\alpha(x_1) \, \Otil_\beta(x_2) } = \frac{\delta_{\alpha \beta}}{\abs{x_{12}}^{2 \Dn_\alpha}} \kla{1 - \lambda \gamma_\alpha \ln \abs{x_{12} \mu}^2 + O(\lambda^2)} 
= \frac{\delta_{\alpha \beta}}{\abs{x_{12}}^{2 \Dn_\alpha} \abs{x_{12} \mu}^{2 \lambda \gamma_\alpha}}\, ,
\end{equation}
allowing one to extract the scheme independent
one-loop scaling dimensions $\Df_\alpha = \Dn_\alpha + \lambda \gamma_\alpha$.

Moving on to three-point functions of the un-renormalized diagonal operators $\cO_{\alpha}$
one obtains to the one-loop order in $\lambda$
\begin{align}
\label{3ptfct}
\langle \, \cO_\alpha(x_1)\, &\cO_\beta(x_2) \, \cO_\gamma(x_3)\,\rangle =
\frac{1}{|x_{12}|^{\Dn_\alpha+\Dn_\beta-\Dn_\gamma}\,|x_{23}|^{\Dn_\beta+\Dn_\gamma-\Dn_\alpha}\,
|x_{31}|^{\Dn_\gamma+\Dn_\alpha-\Dn_\beta} }\, \nn\\
&\times \left [ C^{(0)}_{\alpha\beta\gamma} \left ( 1+ \frac{1}{2}\, \lambda\, \left \{ \gamma_\alpha\,
\ln \frac{\epsilon^2\, x_{23}^2}{x_{12}^2\, x_{31}^2} +  \gamma_\beta\,
\ln \frac{\epsilon^2\, x_{31}^2}{x_{12}^2\, x_{23}^2} +  \gamma_\gamma\,
\ln \frac{\epsilon^2\, x_{12}^2}{x_{23}^2\, x_{31}^2} \,\right \}\,
\right ) + \lambda\,  \tilde C^{(1)}_{\alpha\beta\gamma}\,
 \right ]
\end{align}
Now again the finite one-loop contribution to the structure
constant $\tilde C^{(1)}_{\alpha\beta\gamma}$ is scheme dependent 
\cite{Okuyama:2004bd,Alday:2005nd} as it changes under
$\epsilon\to\epsilon\, e^{c}$ as 
\be
\tilde C^{(1)}_{\alpha\beta\gamma}\to \tilde C^{(1)}_{\alpha\beta\gamma}
+c\, (\gamma_\alpha+\gamma_\beta+\gamma_\gamma)\, C^{(0)}_{\alpha\beta\gamma}\, ,
\qquad \mbox{(no sums on the indices)}\, .
\ee
However, the following combination of the unrenormalized three-point function 
structure constant and the normalization is scheme independent
\be
C^{(1)}_{\alpha\beta\gamma}:= \tilde C^{(1)}_{\alpha\beta\gamma} -\frac{1}{2}\, (
g_{\alpha} \,C^{(0)}_{\alpha\beta\gamma} + g_{\beta} \,C^{(0)}_{\alpha\beta\gamma}
+ g_{\gamma} \,C^{(0)}_{\alpha\beta\gamma}\,
) \, .
\label{schemeindepc}
\ee
This is the only datum to be extracted from three-point functions. It also
directly arises as the structure constant in the three-point function of the
renormalized operators $\Otil_{\alpha}$ 
\be
 \skla{\Otil_\alpha(x_1) \, \Otil_\beta(x_2) \, \Otil_\gamma(x_3)} = \frac{C_{\alpha \beta \gamma}}{\abs{x_{12}}^{\Df_\alpha + \Df_\beta - \Df_\gamma} \abs{x_{23}}^{\Df_\beta + \Df_\gamma - \Df_\alpha} \abs{x_{13}}^{\Df_\alpha + \Df_\gamma - \Df_\beta} \abs{\mu}^{\lambda (\gamma_\alpha + \gamma_\beta + \gamma_\gamma)} } \, ,
\label{3ptconv}
\ee
where 
$
 C_{\alpha \beta \gamma} = C^{(0)}_{\alpha \beta \gamma} + \lambda \,
 C^{(1)}_{\alpha \beta \gamma} + O(\lambda^2)\, 
$
is the scheme independent structure constant of \eqn{schemeindepc}.

An important point is the following. If one wishes to compute the \emph{one-loop} piece
$C^{(1)}_{\alpha\beta\gamma}$ starting from a generic basis of operators 
one has to resolve the mixing problem at the \emph{two-loop} order. This is so as
the resulting mixing matrix $M_{\alpha\beta}$ will then receive $O(\lambda)$ terms which 
will contribute to the final $C^{(1)}_{\alpha\beta\gamma}$ 
through tree-level contractions. If, however,
the degeneracy for a given set of states in a representation of $\alg{psu}(2,2|4)$ has
been completely lifted already at the one-loop order, then this two-loop mixing effect will be absent as
the mixing matrix $M_{\alpha\beta}$ cannot receive further corrections. Looking at short
primary operators this indeed turns out to be the case up for a large class of short
operators, as we will discuss later.

\section{The one-loop planar dressing formulae}
\label{three}

\subsection{Derivation}

In this section we derive an efficient set of combinatorial dressing formulae to 
dress up tree-level graphs to one-loop. Similar formulae appeared in \cite{Drukker:2008pi}.

Following \cite{Beisert:2002bb} we introduce the 4d  propagator and the relevant
one-loop integrals in configuration space
\begin{align}
 I_{12} &= \frac{1}{(2\pi)^2 x_{12}^2}\, ,\nn\\
 Y_{123} &= \int \dv{w} I_{1w} I_{2w} I_{3w}\, ,\nn \\
 X_{1234} &= \int \dv{w} I_{1w} I_{2w} I_{3w} I_{4w}\, ,\nn\\
 H_{12,34} &= \int \dv{v} \dv{w} I_{1v} I_{2v} I_{vw} I_{3w} I_{4w}\, ,\nn \\
 F_{12,34} &= \frac{(\partial_1 - \partial_2) \cdot (\partial_3 - \partial_4) H_{12,34}}{I_{12} I_{34}}\, .
\end{align} 
We have put the space-time points as indices to the function to make the expressions 
more compact. These functions are all finite except in certain limits. For example 
$Y_{123}$ , $X_{1234}$ and $H_{12,34}$ diverge logarithmically when $x_{1}\to x_{2}$. In point 
splitting regularization one has the limiting formulae ($\lim_{i\to j}x^{2}_{ij}=\epsilon^{2}$)
\begin{align}
 X_{1123}  &= -\frac{1}{16 \pi^2}\, I_{12} I_{13} \kla{\ln \frac{x_{23}^2 \e^2}{x_{12}^2 x_{13}^2} -2}, \label{eqn:PointSplittingLimits1:X1123}\\
 Y_{112}   &= -\frac{1}{16 \pi^2}\, I_{12} \kla{\ln \frac{\e^2}{x_{12}^2} -2} = Y_{122},\label{eqn:PointSplittingLimits2:Y112}\\
 F_{12,13} &= -\frac{1}{16 \pi^2} \, \kla{\ln \frac{\e^2}{x_{23}^2} -2} + Y_{123} \kla{ \frac{1}{I_{12}} + \frac{1}{I_{13}} - \frac{2}{I_{23}} },\label{eqn:PointSplittingLimits3:F1213}\\
 X_{1122}  &= -\frac{1}{8 \pi^2} I_{12}^2 \kla{\ln \frac{\e^2}{x_{12}^2} -1},\label{eqn:PointSplittingLimits4:X1122}\\
 F_{12,12} &= -\frac{1}{8 \pi^2} \,\kla{\ln \frac{\e^2}{x_{12}^2} -3}.\label{eqn:PointSplittingLimits5:F1212}
\end{align}

We introduce a graphical symbol for the scalar propagators and work in a normalization
where
\begin{equation}
 \skla{\phi^{I}(x_1) \phi^{J}(x_2)}_{\text{tree}} u_1^{I} u_2^{J} = 
 \begin{minipage}[h]{10mm}\begin{center}\footnotesize

\begin{fmffile}{ScalarPropagator}
\begin{fmfgraph*}(7,7)
\fmfpen{thick}
\fmftop{a1}
\fmfbottom{a2}
\fmfdot{a1,a2}
\fmf{plain}{a1,a2}
\fmflabel{$u_1$}{a1}
\fmflabel{$u_2$}{a2}
\end{fmfgraph*}
\end{fmffile}

\end{center}\end{minipage} = (u_1 \cdot u_2) \, I_{12}\, , \phantom{\Bigg(}
\end{equation}
here the $SO(6)$-indices of the scalar fields are contracted with dummy
six-vectors $u_1^{I}$ and $u_2^{J}$ for bookmarking purposes.

The one-loop corrections are then built of the following three components
\begin{alignat}{2}
\begin{minipage}[h]{18mm}\begin{center}\footnotesize

\begin{fmffile}{LoopSelfEnergy}
\begin{fmfgraph*}(7,7)
\fmfpen{thick}
\fmfleft{a1}
\fmfright{a2}
\fmf{plain}{a1,v,a2}
\fmfdot{a1,a2}
\fmfblob{6}{v}
\fmfdot{a1,a2}
\fmflabel{$u_1$}{a1}
\fmflabel{$u_2$}{a2}
\end{fmfgraph*}
\end{fmffile}

\end{center}\end{minipage} &= - \lambda (u_1 \cdot u_2) \, I_{12} \, \frac{Y_{112} + Y_{122}}{I_{12}}\label{eqn:BasicInteractionsLoopSelfEnergy} &&\quad \quad \text{(self-energy),}\phantom{\Bigg(}\\
\begin{minipage}[h]{18mm}\begin{center}\footnotesize

\begin{fmffile}{LoopGluon}
\begin{fmfgraph*}(7,7)
\fmfpen{thick}
\fmfright{a4,a2}
\fmfleft{a3,a1}
\fmf{plain}{a1,v1,a2}
\fmf{plain}{a3,v2,a4}
\fmffreeze
\fmf{wiggly,width=1}{v1,v2}
\fmfdot{a1,a2,a3,a4}
\fmfv{label=$u_1$,label.dist=1.5}{a1}
\fmfv{label=$u_2$,label.dist=1.5}{a2}
\fmfv{label=$u_3$,label.dist=1.5}{a3}
\fmfv{label=$u_4$,label.dist=1.5}{a4}
\end{fmfgraph*}
\end{fmffile}

\end{center}\end{minipage} &= \frac{\lambda}{2} (u_1 \cdot u_2) (u_3 \cdot u_4) \, I_{12} \, I_{34} \, F_{12,34}\label{eqn:BasicInteractionsLoopGluon} &&\quad \quad \text{(gluon),}\phantom{\Bigg(}\\
\begin{minipage}[h]{18mm}\begin{center}\footnotesize

\begin{fmffile}{LoopVertex}
\begin{fmfgraph*}(7,7)
\fmfpen{thick}
\fmfright{a4,a2}
\fmfleft{a3,a1}
\fmf{plain}{a1,v,a4}
\fmf{plain}{a3,v,a2}
\fmfdot{a1,a2,a3,a4,v}
\fmfv{label=$u_1$,label.dist=1.5}{a1}
\fmfv{label=$u_2$,label.dist=1.5}{a2}
\fmfv{label=$u_3$,label.dist=1.5}{a3}
\fmfv{label=$u_4$,label.dist=1.5}{a4}
\end{fmfgraph*}
\end{fmffile}

\end{center}\end{minipage} &= \frac{\lambda}{2} \big[2 (u_2 \cdot u_3) (u_1 \cdot u_4) - (u_2 \cdot u_4) (u_1 \cdot u_3) \phantom{\Bigg(} \hspace{-5cm} &&\nn\\ &\qquad
 - (u_1 \cdot u_2) (u_3 \cdot u_4)\big] \, X_{1234}\label{eqn:BasicInteractionsLoopVertex} &&\quad \quad  \text{(vertex).}
\end{alignat}
With these basic interactions we can now diagrammatically dress up the tree-level
two- and three-point correlation functions to the one-loop level.
To do so we note that a generic planar three-point function will be made of two-gon and three-gon
sub-graphs which need to be dressed, see figure 1.

\begin{figure}
 \centering
 \raisebox{0.5cm}{\includegraphics[width=4cm]{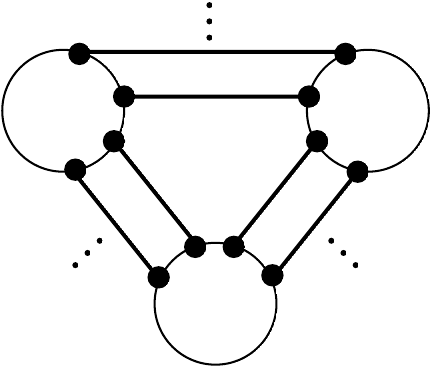}}
 \caption{The generic tree-level three-point function.}
 \label{fig:Ord3pt}
\end{figure}

For the two-gon dressing one finds the basic dressing formula
\begin{align}
\skla{\begin{minipage}[h]{12mm}\begin{center}\footnotesize

\begin{fmffile}{2gon}
\begin{fmfgraph*}(5,14)
\fmfstraight
\fmfleft{dummy1,p1,a2,a1,p2,dummy2}
\fmfright{dummy3,p3,b1,b2,p4,dummy4}
\fmfpen{thin}
\fmf{plain}{dummy1,p1,a2,a1,p2,dummy2}
\fmf{plain}{dummy3,p3,b1,b2,p4,dummy4}
\fmffreeze
\fmf{plain,width=2}{a1,b2}
\fmf{plain,width=2}{a2,b1}
\fmf{dots}{p1,p3}
\fmf{dots}{p2,p4}
\fmfdot{a1,a2,b1,b2}
\fmfv{label=$u_1$,label.dist=1.5thick}{a1}
\fmfv{label=$u_2$,label.dist=1.5thick}{a2}
\fmfv{label=$v_1$,label.dist=1.5thick}{b1}
\fmfv{label=$v_2$,label.dist=1.5thick}{b2}
\fmflabel{$x_1$}{dummy1}
\fmflabel{$x_2$}{dummy3}
\end{fmfgraph*}
\end{fmffile}

\end{center}\end{minipage}}_\text{1-loop} &= \begin{minipage}[h]{10mm}\begin{center}\footnotesize

\begin{fmffile}{2gonVertex}
\begin{fmfgraph*}(5,14)
\fmfstraight
\fmfleft{dummy1,p1,a2,a1,p2,dummy2}
\fmfright{dummy3,p3,b1,b2,p4,dummy4}
\fmfpen{thin}
\fmf{plain}{dummy1,p1,a2,a1,p2,dummy2}
\fmf{plain}{dummy3,p3,b1,b2,p4,dummy4}
\fmffreeze
\fmf{plain,width=2}{a1,v,b1}
\fmf{plain,width=2}{a2,v,b2}
\fmf{dots}{p1,p3}
\fmf{dots}{p2,p4}
\fmfdot{a1,a2,b1,b2,v}
\end{fmfgraph*}
\end{fmffile}

\end{center}\end{minipage} + \begin{minipage}[h]{10mm}\begin{center}\footnotesize

\begin{fmffile}{2gonGluon}
\begin{fmfgraph*}(5,14)
\fmfstraight
\fmfleft{dummy1,p1,a2,a1,p2,dummy2}
\fmfright{dummy3,p3,b1,b2,p4,dummy4}
\fmfpen{thin}
\fmf{plain}{dummy1,p1,a2,a1,p2,dummy2}
\fmf{plain}{dummy3,p3,b1,b2,p4,dummy4}
\fmffreeze
\fmf{plain,width=2}{a1,v1,b2}
\fmf{plain,width=2}{a2,v2,b1}
\fmffreeze
\fmf{wiggly}{v1,v2}
\fmf{dots}{p1,p3}
\fmf{dots}{p2,p4}
\fmfdot{a1,a2,b1,b2}
\end{fmfgraph*}
\end{fmffile}

\end{center}\end{minipage} + \frac{1}{2} \; \begin{minipage}[h]{10mm}\begin{center}\footnotesize

\begin{fmffile}{2gonSelfEnergyTop}
\begin{fmfgraph*}(5,14)
\fmfstraight
\fmfleft{dummy1,p1,a2,a1,p2,dummy2}
\fmfright{dummy3,p3,b1,b2,p4,dummy4}
\fmfpen{thin}
\fmf{plain}{dummy1,p1,a2,a1,p2,dummy2}
\fmf{plain}{dummy3,p3,b1,b2,p4,dummy4}
\fmffreeze
\fmf{plain,width=2}{a1,v,b2}
\fmf{plain,width=2}{a2,b1}
\fmfblob{6}{v}
\fmf{dots}{p1,p3}
\fmf{dots}{p2,p4}
\fmfdot{a1,a2,b1,b2}
\end{fmfgraph*}
\end{fmffile}

\end{center}\end{minipage} + \frac{1}{2} \; \begin{minipage}[h]{10mm}\begin{center}\footnotesize

\begin{fmffile}{2gonSelfEnergyBottom}
\begin{fmfgraph*}(5,14)
\fmfstraight
\fmfleft{dummy1,p1,a2,a1,p2,dummy2}
\fmfright{dummy3,p3,b1,b2,p4,dummy4}
\fmfpen{thin}
\fmf{plain}{dummy1,p1,a2,a1,p2,dummy2}
\fmf{plain}{dummy3,p3,b1,b2,p4,dummy4}
\fmffreeze
\fmf{plain,width=2}{a1,b2}
\fmf{plain,width=2}{a2,v,b1}
\fmfblob{6}{v}
\fmf{dots}{p1,p3}
\fmf{dots}{p2,p4}
\fmfdot{a1,a2,b1,b2}
\end{fmfgraph*}
\end{fmffile}

\end{center}\end{minipage} \nnl
& = I_{12}^{2}\,\, \frac{\lambda}{8\pi^2}\, \Bigl (\ln\frac{\epsilon^2}{x_{12}^2}-1\Bigr )\,
\Bigl ({u_1\cdot v_2\, v_1\cdot u_2}-{u_1\cdot u_2\, v_1\cdot v_2 - \frac{1}{2}\, u_1\cdot v_1\, u_2\cdot v_2 } \Bigr )\,  \nn \\
&=  I_{12}^2 \, \frac{\lambda}{8 \pi^2} \Bigl (\ln \frac{\e^2}{x_{12}^2} - 1\Bigr ) \Bigg( \begin{minipage}[h]{8mm}\begin{center}\footnotesize

\begin{fmffile}{2gonBlank}
\begin{fmfgraph*}(5,14)
\fmfstraight
\fmfleft{dummy1,p1,a2,a1,p2,dummy2}
\fmfright{dummy3,p3,b1,b2,p4,dummy4}
\fmfpen{thin}
\fmf{plain}{dummy1,p1,a2,a1,p2,dummy2}
\fmf{plain}{dummy3,p3,b1,b2,p4,dummy4}
\fmffreeze
\fmf{plain,width=2}{a1,b2}
\fmf{plain,width=2}{a2,b1}
\fmf{dots}{p1,p3}
\fmf{dots}{p2,p4}
\fmfdot{a1,a2,b1,b2}
\end{fmfgraph*}
\end{fmffile}

\end{center}\end{minipage} - \begin{minipage}[h]{8mm}\begin{center}\footnotesize

\begin{fmffile}{2gonCrossed}
\begin{fmfgraph*}(5,14)
\fmfstraight
\fmfleft{dummy1,p1,a2,a1,p2,dummy2}
\fmfright{dummy3,p3,b1,b2,p4,dummy4}
\fmfpen{thin}
\fmf{plain}{dummy1,p1,a2,a1,p2,dummy2}
\fmf{plain}{dummy3,p3,b1,b2,p4,dummy4}
\fmffreeze
\fmf{plain,width=2}{a1,b1}
\fmf{plain,width=2}{a2,b2}
\fmf{dots}{p1,p3}
\fmf{dots}{p2,p4}
\fmfdot{a1,a2,b1,b2}
\end{fmfgraph*}
\end{fmffile}

\end{center}\end{minipage} + \frac{1}{2} \begin{minipage}[h]{8mm}\begin{center}\footnotesize

\begin{fmffile}{2gonSelf}
\begin{fmfgraph*}(5,14)
\fmfstraight
\fmfleft{dummy1,p1,a2,a1,p2,dummy2}
\fmfright{dummy3,p3,b1,b2,p4,dummy4}
\fmfpen{thin}
\fmf{plain}{dummy1,p1,a2,a1,p2,dummy2}
\fmf{plain}{dummy3,p3,b1,b2,p4,dummy4}
\fmfcurved
\fmf{plain,left,width=2,tension=0}{a1,a2}
\fmf{plain,left,width=2,tension=0}{b1,b2}
\fmf{dots}{p1,p3}
\fmf{dots}{p2,p4}
\fmfdot{a1,a2,b1,b2}
\end{fmfgraph*}
\end{fmffile}

\end{center}\end{minipage} \Bigg)\, ,
\label{eqn:2-Gon Dressing}
\end{align}
where the diagrams in the last line only stand for the index contractions not for 
propagators. This contraction structure is of course that of an integrable 
nearest neighbor $SO(6)$ vector spin-chain Hamiltonian as was first noted 
in \cite{Minahan:2002ve}.

Analogously, for the three-gon we find
\begin{align}
\skla{\begin{minipage}[h]{20mm}\begin{center}\footnotesize

\begin{fmffile}{3gon}
\begin{fmfgraph*}(14,14)
\fmfpen{thin}
\fmfsurroundn{d}{6}
\fmffreeze
\fmf{plain}{d1,b1,b2,d2}
\fmf{plain}{d3,a1,a2,d4}
\fmf{plain}{d5,c1,c2,d6}
\fmf{dots}{d2,d3}
\fmf{dots}{d4,d5}
\fmf{dots}{d6,d1}
\fmffreeze
\fmf{plain,width=2}{a1,b2}
\fmf{plain,width=2}{a2,c1}
\fmf{plain,width=2}{b1,c2}
\fmfdot{a1,a2,b1,b2,c1,c2}
\fmfv{label=$u_1$,label.dist=1.5thick}{a1}
\fmfv{label=$u_2$,label.dist=1.5thick}{a2}
\fmfv{label=$v_1$,label.dist=1.5thick}{b1}
\fmfv{label=$v_2$,label.dist=1.5thick}{b2}
\fmfv{label=$w_1~$}{c1}
\fmfv{label=$~w_2$}{c2}
\end{fmfgraph*}
\end{fmffile}

\end{center}\end{minipage}}_\text{1-loop} &=
\frac{1}{2} \; \begin{minipage}[h]{20mm}\begin{center}\footnotesize

\begin{fmffile}{3gonSelfEnergyA}
\begin{fmfgraph*}(14,14)
\fmfpen{thin}
\fmfsurroundn{d}{6}
\fmffreeze
\fmf{plain}{d1,b1,b2,d2}
\fmf{plain}{d3,a1,a2,d4}
\fmf{plain}{d5,c1,c2,d6}
\fmf{dots}{d2,d3}
\fmf{dots}{d4,d5}
\fmf{dots}{d6,d1}
\fmffreeze
\fmf{plain,width=2}{a1,v,b2}
\fmf{plain,width=2}{a2,c1}
\fmf{plain,width=2}{b1,c2}
\fmfdot{a1,a2,b1,b2,c1,c2}
\fmfblob{6}{v}
\fmfv{label=$u_1$,label.dist=1.5thick}{a1}
\fmfv{label=$u_2$,label.dist=1.5thick}{a2}
\fmfv{label=$v_1$,label.dist=1.5thick}{b1}
\fmfv{label=$v_2$,label.dist=1.5thick}{b2}
\fmfv{label=$w_1~$}{c1}
\fmfv{label=$~w_2$}{c2}
\end{fmfgraph*}
\end{fmffile}

\end{center}\end{minipage} + \begin{minipage}[h]{20mm}\begin{center}\footnotesize

\begin{fmffile}{3gonGluonA}
\begin{fmfgraph*}(14,14)
\fmfpen{thin}
\fmfsurroundn{d}{6}
\fmffreeze
\fmf{plain}{d1,b1,b2,d2}
\fmf{plain}{d3,a1,a2,d4}
\fmf{plain}{d5,c1,c2,d6}
\fmf{dots}{d2,d3}
\fmf{dots}{d4,d5}
\fmf{dots}{d6,d1}
\fmffreeze
\fmf{plain,width=2}{a1,v1,b2}
\fmf{plain,width=2}{a2,v2,c1}
\fmf{plain,width=2}{b1,c2}
\fmffreeze
\fmf{wiggly}{v1,v2}
\fmfdot{a1,a2,b1,b2,c1,c2}
\fmfv{label=$u_1$,label.dist=1.5thick}{a1}
\fmfv{label=$u_2$,label.dist=1.5thick}{a2}
\fmfv{label=$v_1$,label.dist=1.5thick}{b1}
\fmfv{label=$v_2$,label.dist=1.5thick}{b2}
\fmfv{label=$w_1~$}{c1}
\fmfv{label=$~w_2$}{c2}
\end{fmfgraph*}
\end{fmffile}

\end{center}\end{minipage} 
+\begin{minipage}[h]{20mm}\begin{center}\footnotesize

\begin{fmffile}{3gonVertexA}
\begin{fmfgraph*}(14,14)
\fmfpen{thin}
\fmfsurroundn{d}{6}
\fmffreeze
\fmf{plain}{d1,b1,b2,d2}
\fmf{plain}{d3,a1,a2,d4}
\fmf{plain}{d5,c1,c2,d6}
\fmf{dots}{d2,d3}
\fmf{dots}{d4,d5}
\fmf{dots}{d6,d1}
\fmffreeze
\fmf{plain,width=2}{a2,v,b2}
\fmf{plain,width=2}{a1,v,c1}
\fmf{plain,width=2}{b1,c2}
\fmffreeze
\fmfdot{a1,a2,b1,b2,c1,c2,v}
\fmfv{label=$u_1$,label.dist=1.5thick}{a1}
\fmfv{label=$u_2$,label.dist=1.5thick}{a2}
\fmfv{label=$v_1$,label.dist=1.5thick}{b1}
\fmfv{label=$v_2$,label.dist=1.5thick}{b2}
\fmfv{label=$w_1~$}{c1}
\fmfv{label=$~w_2$}{c2}
\end{fmfgraph*}
\end{fmffile}

\end{center}\end{minipage} + \mbox{2 permutations} \nn\\
&= I_{12} I_{13} I_{23} \times \frac{\lambda}{16 \pi^2} \nnl
&\times \Bigg[\kla{\lnl{23}{12}{13} - 2} \kla{\begin{minipage}[h]{16mm}\begin{center}\footnotesize

\begin{fmffile}{3gonBlank}
\begin{fmfgraph*}(14,14)
\fmfpen{thin}
\fmfsurroundn{d}{6}
\fmffreeze
\fmf{plain}{d1,b1,b2,d2}
\fmf{plain}{d3,a1,a2,d4}
\fmf{plain}{d5,c1,c2,d6}
\fmf{dots}{d2,d3}
\fmf{dots}{d4,d5}
\fmf{dots}{d6,d1}
\fmffreeze
\fmf{plain,width=2}{a1,b2}
\fmf{plain,width=2}{a2,c1}
\fmf{plain,width=2}{b1,c2}
\fmfdot{a1,a2,b1,b2,c1,c2}
\end{fmfgraph*}
\end{fmffile}

\end{center}\end{minipage} - 
\begin{minipage}[h]{16mm}\begin{center}\footnotesize

\begin{fmffile}{3gonCrossedA}
\begin{fmfgraph*}(14,14)
\fmfpen{thin}
\fmfsurroundn{d}{6}
\fmffreeze
\fmf{plain}{d1,b1,b2,d2}
\fmf{plain}{d3,a1,a2,d4}
\fmf{plain}{d5,c1,c2,d6}
\fmf{dots}{d2,d3}
\fmf{dots}{d4,d5}
\fmf{dots}{d6,d1}
\fmffreeze
\fmf{plain,width=2}{a2,b2}
\fmf{plain,width=2}{a1,c1}
\fmf{plain,width=2}{b1,c2}
\fmfdot{a1,a2,b1,b2,c1,c2}
\end{fmfgraph*}
\end{fmffile}

\end{center}\end{minipage} + \frac{1}{2} \begin{minipage}[h]{16mm}\begin{center}\footnotesize

\begin{fmffile}{3gonSelfA}
\begin{fmfgraph*}(14,14)
\fmfpen{thin}
\fmfsurroundn{d}{6}
\fmffreeze
\fmf{plain}{d1,b1,b2,d2}
\fmf{plain}{d3,a1,a2,d4}
\fmf{plain}{d5,c1,c2,d6}
\fmf{dots}{d2,d3}
\fmf{dots}{d4,d5}
\fmf{dots}{d6,d1}
\fmffreeze
\fmfcurved
\fmf{plain,left,width=2}{a1,a2}
\fmf{plain,right,width=2}{b2,c1}
\fmf{plain,width=2}{b1,c2}
\fmfdot{a1,a2,b1,b2,c1,c2}
\end{fmfgraph*}
\end{fmffile}

\end{center}\end{minipage}} \nnl
&+ \kla{\lnl{13}{12}{23} - 2} \kla{\begin{minipage}[h]{16mm}\begin{center}\footnotesize

\begin{fmffile}{3gonBlank}
\begin{fmfgraph*}(14,14)
\fmfpen{thin}
\fmfsurroundn{d}{6}
\fmffreeze
\fmf{plain}{d1,b1,b2,d2}
\fmf{plain}{d3,a1,a2,d4}
\fmf{plain}{d5,c1,c2,d6}
\fmf{dots}{d2,d3}
\fmf{dots}{d4,d5}
\fmf{dots}{d6,d1}
\fmffreeze
\fmf{plain,width=2}{a1,b2}
\fmf{plain,width=2}{a2,c1}
\fmf{plain,width=2}{b1,c2}
\fmfdot{a1,a2,b1,b2,c1,c2}
\end{fmfgraph*}
\end{fmffile}

\end{center}\end{minipage} 
- \begin{minipage}[h]{16mm}\begin{center}\footnotesize

\begin{fmffile}{3gonCrossedB}
\begin{fmfgraph*}(14,14)
\fmfpen{thin}
\fmfsurroundn{d}{6}
\fmffreeze
\fmf{plain}{d1,b1,b2,d2}
\fmf{plain}{d3,a1,a2,d4}
\fmf{plain}{d5,c1,c2,d6}
\fmf{dots}{d2,d3}
\fmf{dots}{d4,d5}
\fmf{dots}{d6,d1}
\fmffreeze
\fmf{plain,width=2}{a1,b1}
\fmf{plain,width=2}{a2,c1}
\fmf{plain,width=2}{b2,c2}
\fmfdot{a1,a2,b1,b2,c1,c2}
\end{fmfgraph*}
\end{fmffile}

\end{center}\end{minipage} + \frac{1}{2} \begin{minipage}[h]{16mm}\begin{center}\footnotesize

\begin{fmffile}{3gonSelfB}
\begin{fmfgraph*}(14,14)
\fmfpen{thin}
\fmfsurroundn{d}{6}
\fmffreeze
\fmf{plain}{d1,b1,b2,d2}
\fmf{plain}{d3,a1,a2,d4}
\fmf{plain}{d5,c1,c2,d6}
\fmf{dots}{d2,d3}
\fmf{dots}{d4,d5}
\fmf{dots}{d6,d1}
\fmffreeze
\fmf{plain,left,width=2}{a1,c2}
\fmf{plain,width=2}{a2,c1}
\fmf{plain,left,width=2}{b1,b2}
\fmfdot{a1,a2,b1,b2,c1,c2}
\end{fmfgraph*}
\end{fmffile}

\end{center}\end{minipage}} \nnl
&+ \kla{\lnl{12}{13}{23} - 2} \kla{\begin{minipage}[h]{16mm}\begin{center}\footnotesize

\begin{fmffile}{3gonBlank}
\begin{fmfgraph*}(14,14)
\fmfpen{thin}
\fmfsurroundn{d}{6}
\fmffreeze
\fmf{plain}{d1,b1,b2,d2}
\fmf{plain}{d3,a1,a2,d4}
\fmf{plain}{d5,c1,c2,d6}
\fmf{dots}{d2,d3}
\fmf{dots}{d4,d5}
\fmf{dots}{d6,d1}
\fmffreeze
\fmf{plain,width=2}{a1,b2}
\fmf{plain,width=2}{a2,c1}
\fmf{plain,width=2}{b1,c2}
\fmfdot{a1,a2,b1,b2,c1,c2}
\end{fmfgraph*}
\end{fmffile}

\end{center}\end{minipage} - 
\begin{minipage}[h]{16mm}\begin{center}\footnotesize

\begin{fmffile}{3gonCrossedC}
\begin{fmfgraph*}(14,14)
\fmfpen{thin}
\fmfsurroundn{d}{6}
\fmffreeze
\fmf{plain}{d1,b1,b2,d2}
\fmf{plain}{d3,a1,a2,d4}
\fmf{plain}{d5,c1,c2,d6}
\fmf{dots}{d2,d3}
\fmf{dots}{d4,d5}
\fmf{dots}{d6,d1}
\fmffreeze
\fmf{plain,width=2}{a1,b2}
\fmf{plain,width=2}{a2,c2}
\fmf{plain,width=2}{b1,c1}
\fmfdot{a1,a2,b1,b2,c1,c2}
\end{fmfgraph*}
\end{fmffile}

\end{center}\end{minipage} + \frac{1}{2} \begin{minipage}[h]{16mm}\begin{center}\footnotesize

\begin{fmffile}{3gonSelfC}
\begin{fmfgraph*}(14,14)
\fmfpen{thin}
\fmfsurroundn{d}{6}
\fmffreeze
\fmf{plain}{d1,b1,b2,d2}
\fmf{plain}{d3,a1,a2,d4}
\fmf{plain}{d5,c1,c2,d6}
\fmf{dots}{d2,d3}
\fmf{dots}{d4,d5}
\fmf{dots}{d6,d1}
\fmffreeze
\fmf{plain,width=2}{a1,b2}
\fmf{plain,right,width=2}{a2,b1}
\fmf{plain,left,width=2}{c1,c2}
\fmfdot{a1,a2,b1,b2,c1,c2}
\end{fmfgraph*}
\end{fmffile}

\end{center}\end{minipage}} \Bigg].
\label{eqn:3-Gon Dressing}
\end{align}
Again the graphs in the last three lines only represent the index contractions.
Interestingly a similar structure to the integrable spin-chain Hamiltonian
of \eqn{eqn:2-Gon Dressing} emerges also for the one-loop three-gon interactions.

\subsection{Gauge invariance and Wilson line contributions}

There is one important point we have not addressed so far. The point splitting regularization method
that we employed violates gauge invariance as the space-time locations of the
two neighboring operators in the trace are no longer coincident. The natural way to recover 
gauge invariance is to connect the two split points through a straight Wilson line. This, however, gives rise
to new diagrams not yet accounted for in which a gluon is radiated off the Wilson line.  
Luckily we are able to show that this
contribution vanishes entirely at the one-loop level for $|\epsilon|\to 0$.

Setting $\epsilon^{\mu}=x^{\mu}_{13}$
the Wilson line is parametrized by
\be
x^{\mu}(\tau) = x_{3}^{\mu}+\epsilon^{\mu}\, \tau\, , \quad \tau\in[0,1]\, .
\ee
We then have the contribution
\begin{align}
\begin{minipage}[h]{20mm}\begin{center}\footnotesize

\begin{fmffile}{WilsonLine}
\begin{fmfgraph*}(16,16)
\fmfstraight
\fmfpen{thick}
\fmfright{d4,a4,tau2,a2,d2}
\fmfleft{d3,a3,tau,a1,d1}
\fmf{plain}{a1,v1,a2}
\fmf{plain}{a3,v2,a4}
\fmffreeze
\fmf{dbl_plain,width=1}{a3,tau,a1}
\fmf{wiggly,width=1}{v1,tau}
\fmf{dots}{d1,a1}
\fmf{dots}{d2,d4}
\fmf{dots}{d3,a3}
\fmfdot{a1,a2,a3,a4,tau,v1}
\fmfv{label=$1$}{a1}
\fmfv{label=$2$}{a2}
\fmfv{label=$3$}{a3}
\fmfv{label=$4$}{a4}
\fmfv{label=$\tau$}{tau}
\fmfv{label=$\omega$}{v1}
\end{fmfgraph*}
\end{fmffile}

\end{center}\end{minipage} \,\,\, &=
\lambda (u_{1}\cdot u_{2})(u_{3}\cdot u_{4})\, 
\int_{0}^{1}d\tau \, \epsilon\cdot(\partial_{1}-\partial_{2})\, Y_{12\tau} \nn\\
&=
-\frac{2\lambda (u_{1}\cdot u_{2})(u_{3}\cdot u_{4})}{(2\pi)^{6}}\,
\int_{0}^{1}d\tau \int d^{4}\omega \, \frac{\epsilon\cdot x_{1\omega}}
{(x_{1\omega}^{2})^{2}\, x_{2\omega}^{2}\, x_{\tau\omega}^{2}}\, .
\end{align}
This five dimensional integral is by power-counting logarithmically divergent for
coincident points $x_{3},x(\tau)\to x_{1}$ i.e.~$|\epsilon|\to 0$ and one has
\be
\lim_{|\epsilon|\to 0}\int_{0}^{1} d\tau\int d^{4}\omega \, \frac{\epsilon\cdot x_{1\omega}}
{(x_{1\omega}^{2})^{2}\, x_{2\omega}^{2}\, x_{\tau\omega}^{2}} 
\sim \lim_{|\epsilon|\to 0}\, \epsilon\cdot x_{12}\, \left ( 
\ln\frac{\epsilon^{2}}{x_{12}^{2}} +\text{finite} + O(\epsilon)\, \right )
\to 0\, .
\ee
There is also a novel ladder-diagram in which a gluon is exchanged between two Wilson lines
extending from $x_{1}$ to $x_{3}$ and from $x_{2}$ to $x_{4}$. This ladder-graph is 
manifestly finite and vanishes
as $\epsilon^{2}$.
Therefore all the Wilson line contributions to the point splitting regularization vanish at this
order of perturbation theory.

\subsection{Extremal three-point functions}

\begin{figure}
 \centering
 \raisebox{0.5cm}{\includegraphics[width=4cm]{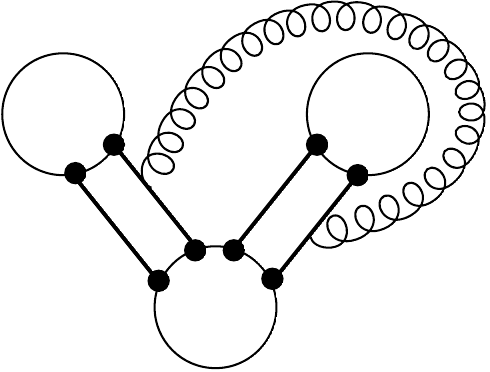}}
 \caption{Additional Feynman-Graphs for extremal three-point functions.}
 \label{fig:Extremal3pt}
\end{figure}

Three-point functions of operators with lengths $\Dn_\alpha$, $\Dn_\beta$ and 
$\Dn_\gamma$ where
$\Dn_\alpha + \Dn_\beta = \Dn_\gamma$ are called extremal. For these extremal functions the dressing formulae above do not hold any longer for two reasons: First, there appear additional diagrams with a gluon exchange or a vertex between non-neighboring propagators as the one in figure \ref{fig:Extremal3pt}. These non-nearest neighbor interactions lead to additional terms in the dressing formulae. Second, unlike non-extremal ones extremal three-point functions with double-trace operators contain the same factor of $N$ as those with single-trace operators. This results in an operator mixing \index{operator!mixing} of single-trace with double-trace operators already at tree-level. This is described in detail in \cite{D'Hoker:1999ea,Okuyama:2004bd}.

We will refrain from studying these extremal three-point correlators in the following. 
In any case the one-loop structure constants follow a simple pattern:
They are a given by a linear function of the anomalous scaling dimensions
of the operators involved \cite{Okuyama:2004bd}
\begin{equation}
 C^{(1)}_{\alpha \beta \gamma, \,\text{extremal}} = \frac{1}{2} \, 
 C^{(0)}_{\alpha \beta \gamma, \, \text{extremal}} \, 
 \kla{\gamma_\alpha + \gamma_\beta - \gamma_\gamma}\, ,
 \label{eqn:form of structure constants for extremal correlators}
\end{equation}
hence the three-point problem has been reduced to the two-point one. In particular
structure constants of protected operators are free of radiative corrections.

\subsection{Two convenient regularization schemes}

We have seen in \eqn{schemeindepc} how to extract the regularization scheme independent
structure constant from a combination of the bare structure constant and the one-loop
finite normalization shifts. As the latter arises from the finite contribution to
the two-gon dressing \eqn{eqn:2-Gon Dressing} one may pick a regularization to simply 
cancel these contributions. I.e.~making the transformation on the point-splitting
parameter
\be
\epsilon \to \sqrt{e}\, \epsilon
\ee
transforms
\be
\lnk{ij}-1 \to \lnk{ij}\, , \qquad \text{and}\qquad
\lnl{ij}{ik}{jk}-2  \to \lnl{ij}{ik}{jk}-1\ .
\ee
Hence in this scheme the finite part of the two-gon dressing vanishes resulting in 
a vanishing finite correction to the two-point functions
\be
g_{\alpha}=0\, ,
\ee
which in turn implies that the bare and the renormalized structure functions
coincide in this scheme
\be
\widetilde{C}^{(1)}_{\alpha \beta \gamma} = C^{(1)}_{\alpha \beta \gamma}\, .
\ee
This implies that the structure function may be read off solely from the three-gon
dressings of the non-extremal correlator, which may be graphically represented by
\begin{align}
 C_{\alpha \beta \gamma}^{(1)} &= -\frac{1}{16 \pi^2} \, \sum_{\substack{\text{cyclic}\\\text{perm.}}} \, \Bigg[ 3 \times \begin{minipage}[h]{16mm}\begin{center}\footnotesize

\begin{fmffile}{3gonBlank}
\begin{fmfgraph*}(14,14)
\fmfpen{thin}
\fmfsurroundn{d}{6}
\fmffreeze
\fmf{plain}{d1,b1,b2,d2}
\fmf{plain}{d3,a1,a2,d4}
\fmf{plain}{d5,c1,c2,d6}
\fmf{dots}{d2,d3}
\fmf{dots}{d4,d5}
\fmf{dots}{d6,d1}
\fmffreeze
\fmf{plain,width=2}{a1,b2}
\fmf{plain,width=2}{a2,c1}
\fmf{plain,width=2}{b1,c2}
\fmfdot{a1,a2,b1,b2,c1,c2}
\end{fmfgraph*}
\end{fmffile}

\end{center}\end{minipage} - \begin{minipage}[h]{16mm}\begin{center}\footnotesize

\begin{fmffile}{3gonCrossedA}
\begin{fmfgraph*}(14,14)
\fmfpen{thin}
\fmfsurroundn{d}{6}
\fmffreeze
\fmf{plain}{d1,b1,b2,d2}
\fmf{plain}{d3,a1,a2,d4}
\fmf{plain}{d5,c1,c2,d6}
\fmf{dots}{d2,d3}
\fmf{dots}{d4,d5}
\fmf{dots}{d6,d1}
\fmffreeze
\fmf{plain,width=2}{a2,b2}
\fmf{plain,width=2}{a1,c1}
\fmf{plain,width=2}{b1,c2}
\fmfdot{a1,a2,b1,b2,c1,c2}
\end{fmfgraph*}
\end{fmffile}

\end{center}\end{minipage} 
 + \frac{1}{2} \times \begin{minipage}[h]{16mm}\begin{center}\footnotesize

\begin{fmffile}{3gonSelfA}
\begin{fmfgraph*}(14,14)
\fmfpen{thin}
\fmfsurroundn{d}{6}
\fmffreeze
\fmf{plain}{d1,b1,b2,d2}
\fmf{plain}{d3,a1,a2,d4}
\fmf{plain}{d5,c1,c2,d6}
\fmf{dots}{d2,d3}
\fmf{dots}{d4,d5}
\fmf{dots}{d6,d1}
\fmffreeze
\fmfcurved
\fmf{plain,left,width=2}{a1,a2}
\fmf{plain,right,width=2}{b2,c1}
\fmf{plain,width=2}{b1,c2}
\fmfdot{a1,a2,b1,b2,c1,c2}
\end{fmfgraph*}
\end{fmffile}

\end{center}\end{minipage} \nnl
 &\bleq - \begin{minipage}[h]{16mm}\begin{center}\footnotesize

\begin{fmffile}{3gonCrossedB}
\begin{fmfgraph*}(14,14)
\fmfpen{thin}
\fmfsurroundn{d}{6}
\fmffreeze
\fmf{plain}{d1,b1,b2,d2}
\fmf{plain}{d3,a1,a2,d4}
\fmf{plain}{d5,c1,c2,d6}
\fmf{dots}{d2,d3}
\fmf{dots}{d4,d5}
\fmf{dots}{d6,d1}
\fmffreeze
\fmf{plain,width=2}{a1,b1}
\fmf{plain,width=2}{a2,c1}
\fmf{plain,width=2}{b2,c2}
\fmfdot{a1,a2,b1,b2,c1,c2}
\end{fmfgraph*}
\end{fmffile}

\end{center}\end{minipage} + \frac{1}{2} \times \begin{minipage}[h]{16mm}\begin{center}\footnotesize

\begin{fmffile}{3gonSelfB}
\begin{fmfgraph*}(14,14)
\fmfpen{thin}
\fmfsurroundn{d}{6}
\fmffreeze
\fmf{plain}{d1,b1,b2,d2}
\fmf{plain}{d3,a1,a2,d4}
\fmf{plain}{d5,c1,c2,d6}
\fmf{dots}{d2,d3}
\fmf{dots}{d4,d5}
\fmf{dots}{d6,d1}
\fmffreeze
\fmf{plain,left,width=2}{a1,c2}
\fmf{plain,width=2}{a2,c1}
\fmf{plain,left,width=2}{b1,b2}
\fmfdot{a1,a2,b1,b2,c1,c2}
\end{fmfgraph*}
\end{fmffile}

\end{center}\end{minipage} 
 - \begin{minipage}[h]{16mm}\begin{center}\footnotesize

\begin{fmffile}{3gonCrossedC}
\begin{fmfgraph*}(14,14)
\fmfpen{thin}
\fmfsurroundn{d}{6}
\fmffreeze
\fmf{plain}{d1,b1,b2,d2}
\fmf{plain}{d3,a1,a2,d4}
\fmf{plain}{d5,c1,c2,d6}
\fmf{dots}{d2,d3}
\fmf{dots}{d4,d5}
\fmf{dots}{d6,d1}
\fmffreeze
\fmf{plain,width=2}{a1,b2}
\fmf{plain,width=2}{a2,c2}
\fmf{plain,width=2}{b1,c1}
\fmfdot{a1,a2,b1,b2,c1,c2}
\end{fmfgraph*}
\end{fmffile}

\end{center}\end{minipage} + \frac{1}{2} \times \begin{minipage}[h]{16mm}\begin{center}\footnotesize

\begin{fmffile}{3gonSelfC}
\begin{fmfgraph*}(14,14)
\fmfpen{thin}
\fmfsurroundn{d}{6}
\fmffreeze
\fmf{plain}{d1,b1,b2,d2}
\fmf{plain}{d3,a1,a2,d4}
\fmf{plain}{d5,c1,c2,d6}
\fmf{dots}{d2,d3}
\fmf{dots}{d4,d5}
\fmf{dots}{d6,d1}
\fmffreeze
\fmf{plain,width=2}{a1,b2}
\fmf{plain,right,width=2}{a2,b1}
\fmf{plain,left,width=2}{c1,c2}
\fmfdot{a1,a2,b1,b2,c1,c2}
\end{fmfgraph*}
\end{fmffile}

\end{center}\end{minipage} \Bigg]\, .
\label{eqn:Structure Constant as 3-Gon Dressing}
\end{align}
Alternatively one may apply the transformation  
\be
\epsilon \to e\, \epsilon
\ee
yielding
\be
\lnk{ij}-1 \to \lnk{ij} +1\, , \qquad \text{and}\qquad
\lnl{ij}{ik}{jk}-2  \to \lnl{ij}{ik}{jk}\, .
\label{here}
\ee

Now the finite contributions to the three-gon dressings vanish and the bare structure
constant may be computed from only dressing the two-gons in the tree-level
correlator
\be
 \widetilde{C}^{(1)}_{\alpha \beta \gamma} = \frac{1}{8 \pi^2} \sum_{\substack{\text{cyclic}\\\text{perm.}}} \; \sum_{\substack{\text{all}\\\text{2-gons}}} \Bigg( \begin{minipage}[h]{8mm}\begin{center}\footnotesize

\begin{fmffile}{2gonBlank}
\begin{fmfgraph*}(5,14)
\fmfstraight
\fmfleft{dummy1,p1,a2,a1,p2,dummy2}
\fmfright{dummy3,p3,b1,b2,p4,dummy4}
\fmfpen{thin}
\fmf{plain}{dummy1,p1,a2,a1,p2,dummy2}
\fmf{plain}{dummy3,p3,b1,b2,p4,dummy4}
\fmffreeze
\fmf{plain,width=2}{a1,b2}
\fmf{plain,width=2}{a2,b1}
\fmf{dots}{p1,p3}
\fmf{dots}{p2,p4}
\fmfdot{a1,a2,b1,b2}
\end{fmfgraph*}
\end{fmffile}

\end{center}\end{minipage} - \begin{minipage}[h]{8mm}\begin{center}\footnotesize

\begin{fmffile}{2gonCrossed}
\begin{fmfgraph*}(5,14)
\fmfstraight
\fmfleft{dummy1,p1,a2,a1,p2,dummy2}
\fmfright{dummy3,p3,b1,b2,p4,dummy4}
\fmfpen{thin}
\fmf{plain}{dummy1,p1,a2,a1,p2,dummy2}
\fmf{plain}{dummy3,p3,b1,b2,p4,dummy4}
\fmffreeze
\fmf{plain,width=2}{a1,b1}
\fmf{plain,width=2}{a2,b2}
\fmf{dots}{p1,p3}
\fmf{dots}{p2,p4}
\fmfdot{a1,a2,b1,b2}
\end{fmfgraph*}
\end{fmffile}

\end{center}\end{minipage} + \frac{1}{2} \begin{minipage}[h]{8mm}\begin{center}\footnotesize

\begin{fmffile}{2gonSelf}
\begin{fmfgraph*}(5,14)
\fmfstraight
\fmfleft{dummy1,p1,a2,a1,p2,dummy2}
\fmfright{dummy3,p3,b1,b2,p4,dummy4}
\fmfpen{thin}
\fmf{plain}{dummy1,p1,a2,a1,p2,dummy2}
\fmf{plain}{dummy3,p3,b1,b2,p4,dummy4}
\fmfcurved
\fmf{plain,left,width=2,tension=0}{a1,a2}
\fmf{plain,left,width=2,tension=0}{b1,b2}
\fmf{dots}{p1,p3}
\fmf{dots}{p2,p4}
\fmfdot{a1,a2,b1,b2}
\end{fmfgraph*}
\end{fmffile}

\end{center}\end{minipage} \Bigg)\, .
 \ee
The scheme independent structure constants can then be calculated using 
 \eqn{schemeindepc} with $g_\alpha=\gamma_\alpha$ by virtue of \eqn{here}, i.e.
\be
 C^{(1)}_{\alpha \beta \gamma} = \widetilde{C}^{(1)}_{\alpha \beta \gamma} - \frac{1}{2} \, C^{(0)}_{\alpha \beta \gamma} \, \kla{\gamma_\alpha  + \gamma_\beta + \gamma_\gamma}. \label{eqn:structure constants as 2-gon dressing}
\ee
In our actual computations we have used both schemes depending on the problem at hand.

\begin{figure}[h]
\centering
 \begin{tabular}{|l|l|l|l|l|l|} \hline
Length & Class & $SU(4)^{\text{parity}}_{\text{length}}$ Rep. & Dim.  & $8\pi^{2}\,\gamma$ & Operator  \\ \hline \hline
 \multirow{2}{*}{\raisebox{-6pt}{2}} & 2A &\ocx{$[0,0,0]^{+}_{2}$} & \ocx{1} &
  \ocx{$6$} & \ocx{$\mathcal{K}$}
 \psx \cline{2-6}
& \ocx{2B} & \ocx{$[0,2,0]^{+}_{2}$} & \ocx{20} & \ocx{$0$} & \ocx{CPO} \psx \hline \hline
 \multirow{2}{*}{\raisebox{-10pt}{3}} 
& \ocx{3B} & \ocx{$[0,1,0]^{-}_{3}$} & \ocx{6} & \ocx{$4$} & \ocx{$\mathcal{O}^{J=1}_{n=1}$} \psx \cline{2-6}
& \ocx{3C}& \ocx{$[0,3,0]^{-}_{3}$} & \ocx{50} & \ocx{$0$} & \ocx{CPO} \psx \hline \hline
 \multirow{6}{*}{\raisebox{-26pt}{4}} & \ocx{4A} & \ocx{$[0,0,0]^{+}_{4}$} & \ocx{1} & 
 \ocx{$\half \, (13 + \sqrt{41})$} & \ocx{$\ast$} \psx \cline{2-6}
& \ocx{4E} & \ocx{$[0,0,0]^{+}_{4}$} & \ocx{1} & \ocx{$\half \, (13 - \sqrt{41})$} & 
\ocx{$\ast$} \psx \cline{2-6}
   & \ocx{4B} & \ocx{$[0,2,0]^{+}_{4}$} & \ocx{20} & \ocx{$5 + \sqrt{5}$} 
   &  \ocx{$\mathcal{O}^{J=2}_{n=2}$} \psx \cline{2-6}
 &\ocx{4F} & \ocx{$[0,2,0]^{+}_{4}$} & \ocx{20} & \ocx{$5 - \sqrt{5}$} & 
 \ocx{$\mathcal{O}^{J=2}_{n=1}$}  \psx \cline{2-6}
 & \ocx{\bf 4C} & \ocx{$[2,0,2]_{4}+[1,0,1]^{-}_{4}$} & \ocx{84 + 15} & \ocx{$6$} & 
\psx \cline{2-6}
& \ocx{4G} & 
 \ocx{$[0,4,0]^{+}_{4}$} & \ocx{105} & \ocx{$0$} & \ocx{CPO} \psx \hline \hline
\multirow{8}{*}{\raisebox{-42pt}{5}} 
& \ocx{\bf 5A} & \ocx{$[0,0,2]^{+}_{5}+[2,0,0]^{+}_{5}$} & \ocx{10 + $\bar{10}$} & \ocx{$7 + \sqrt{13}$} & \psx \cline{2-6}
& \ocx{\bf 5H} & \ocx{$[0,0,2]^{+}_{5}+[2,0,0]^{+}_{5}$} & \ocx{10 + $\bar{10}$} & \ocx{$7 - \sqrt{13}$} & \psx \cline{2-6}
& \ocx{\bf 5D} & \ocx{$[0,1,0]^{-}_{5}$ + {desc}} & \ocx{6 + 252} & \ocx{$5 + \sqrt{5}$} & \ocx{} \psx \cline{2-6}
& \ocx{\bf 5I} & \ocx{$[0,1,0]^{-}_{5}$ + {desc}} & \ocx{6 + 252} & \ocx{$5 - \sqrt{5}$} & \ocx{} \psx \cline{2-6}
& \ocx{\bf 5F} & \ocx{$[1,1,1]^{+}_{5}$ + $[1,1,1]^{-}_{5}$} &\ocx{64 + 64} & \ocx{$5$} & \psx \cline{2-6}
& \ocx{5J} & \ocx{$[0,3,0]^{-}_{5}$} &\ocx{50} & \ocx{$2$} & \ocx{} \psx \cline{2-6}
& \ocx{\bf 5E} & \ocx{$[0,3,0]^{-}_{5}$ + {desc}} &\ocx{50 + 140} & \ocx{$6$} & 
\ocx{$\mathcal{O}^{J=3}_{n=1}$}  \psx \cline{2-6}
& \ocx{5K} & \ocx{$[0,5,0]^{-}_{5}$} &\ocx{196} & \ocx{$0$} & \ocx{CPO} \psx
\cline{2-6}
& \ocx{\bf 5B} & \ocx{$[0,1,0]^{-}_{5}$} &\ocx{\bf 6+6} & \ocx{$10$} & \psx \cline{2-5} \hline
\end{tabular}
\caption{List of all scalar conformal primary operator up to length 5 with their one-loop
anomalous dimensions. Degenerate classes of 
operators are printed in bold-face. $\mathcal{K}$ denotes the Konishi and
 $CPO$ chiral primary operators. The $\mathcal{O}^{J}_{n}$ refer to the BMN singlet
 operators in the nomenclature of \cite{Beisert:2002tn,Georgiou:2009tp} where the quantum
 mixing with fermion and derivative insertions is resolved. The asterix refers to not
 resolved fermion and derivative mixings. 
}
\label{fig:sd}
\end{figure}

\section{Results}

Using the dressing formulae of section \ref{three} one can in principle straightforwardly compute
arbitrary three-point functions by combinatorial means. Clearly, due to the need to sum over
all permutations in these dressing formulae the complexity in the computations grows fast and needs 
to be done on a computer. This has been implemented in a two step procedure. Starting with an
arbitrarily chosen basis of operators all two-point functions are computed and then diagonalized.
Similarly all three-point functions are computed in the original basis 
and then projected to the diagonal basis where
the structure constants can be extracted. For operators up to length three this was done algebraically
with a Mathematica program. Starting with length four the mixing matrix diagonalization could not
be performed algebraically any longer and we had to resort to numerics using Matlab. 
Once the diagonal basis was constructed the numerically obtained structure constants could 
in most cases be again fitted 
to algebraic expressions derived by the algebraic form of the one-loop scaling dimensions. This
could be done in 62 out of 70 cases.


\newcommand{\gmZB}{0} 
\newcommand{\gmDB}{4} 
\newcommand{\gmDC}{0} 
\newcommand{\gmVA}{$\half \, (13 + \sqrt{41})$} 
\newcommand{\gmVB}{$5 + \sqrt{5}$} 
\newcommand{\gmVE}{$\half \, (13 - \sqrt{41})$} 
\newcommand{\gmVF}{$5 - \sqrt{5}$} 
\newcommand{\gmVG}{0} 
\newcommand{\gmFJ}{2} 
\newcommand{\gmFK}{0} 


\subsection{Short primary scalar
operators up to length 5}

We first list all the scalar conformal primary operators up to and including length
five. We have independently constructed this list by an explicit diagonalization 
of the corresponding two-point functions finding complete agreement with the previous
analysis of Beisert \cite{Beisert:2003te,Beisert:2004ry} see figure \ref{fig:sd}. 
Note that there remain degeneracies in the anomalous scaling dimensions $\gamma$
which we indicate in the table through bold face letters.

The operators up to length three and the length four singlets can be explicitly given and read
\begin{align}
 \Op_{2A} &= \sum_{i=1}^6 \tr{\phi^i \phi^i} = \mathcal{K} \\
 \Op_{2B,(ij)} &= \tr{\phi^i \phi^j} & (i<j) \\
 \Op_{2B,i} &= \tr{\phi^i \phi^i} - \frac{1}{\sqrt{3}} \; \mathcal{K} & (i = 2 \dots 6)\\
 \Op_{3B,i} &= \sum_{j=1}^6 \tr{\phi^i \phi^j \phi^j} \\
 \Op_{3C,i(jk)} &= \tr{\phi^i \phi^{(j} \phi^{k)}} & (i<j<k) \\
 \Op_{3C,ij} &= 8 \; \tr{\phi^i \phi^j \phi^j} - \sum_{k=1}^6 \tr{\phi^i \phi^k \phi^k} & (i \not= j, j= 2 \dots 6) \\
 \Op_{3C,i} &= 8 \; \tr{\phi^i \phi^i \phi^i} - 3 \sum_{j=1}^6 \tr{\phi^i \phi^j \phi^j} & (i = 2 \dots 6)\\
 \Op_{4A} &= \sum_{i=1}^6 \sum_{j=1}^6 \ekla{ 4 \; \tr{\phi^i \phi^i \phi^j \phi^j} + \kla{5-\sqrt{41}} \tr{\phi^i \phi^j \phi^i \phi^j} } +\ldots\hspace{-15cm}\\
 \Op_{4E} &= \sum_{i=1}^6 \sum_{j=1}^6 \ekla{ 4 \; \tr{\phi^i \phi^i \phi^j \phi^j} + \kla{5+\sqrt{41}} \tr{\phi^i \phi^j \phi^i \phi^j} } +\ldots\, , \hspace{-15cm}\label{eqn:Length four singlets}
\end{align}
The dots in the last two operators indicate possible operator mixings
with fermion and derivative insertions which have not been resolved so far. Similarly
the operators 4B, 4F and 5E mix with such terms and have been displayed in 
\cite{Georgiou:2008vk,Georgiou:2009tp}\footnote{We thank the authors of this work for important 
discussions on this point.}.  

Below  we list our main results. We computed almost all one-loop structure constants 
for the non-degenerate operators of up to length five of figure 3. 
Note that only three-point functions which do not vanish at
tree-level are listed. We also stress that the majority of results for the fractions 
$C_{\alpha \beta \gamma}^{(1)}/C_{\alpha \beta \gamma}^{(0)}$ have been obtained numerically
and the quoted analytical results represents a biases fit allowing as non-rational factors only the
square root term appearing in the anomalous scaling dimensions of the operators involved in the
patricular three-point function. The numerical
precision in theses fits is typically of order $10^{-5}$ or better, for the raw data see 
the appendix  A.2 of \cite{DiplomarbeitGrossardt}. Finally, the analytically obtained results are highlighted
in bold-face letters.

\begin{center}
 \begin{longtable}{|l|l|l||l|l|l||l|} \hl
 $\Op_\alpha$ & $\Op_\beta$ & $\Op_\gamma$ & $8 \pi^2 \gamma_\alpha$ & $8 \pi^2 \gamma_\beta$ & $8 \pi^2 \gamma_\gamma$ & $-16 \pi^2 C_{\alpha \beta \gamma}^{(1)}/C_{\alpha \beta \gamma}^{(0)}$ \ps\\ \hld \endhead

 \oc{2B} & \oc{3B} & \oc{3B} & \oc{\gmZB} & \oc{\gmDB} & \oc{\gmDB} & \oc{${\bf \frac{8}{3}}$ } \ps\\ \hl
 \oc{2B} & \oc{3B} & \oc{3C} & \oc{\gmZB} & \oc{\gmDB} & \oc{\gmDC} & \oc{{\bf 4}} \ps\\ \hl
 \oc{2B} & \oc{3C} & \oc{3C} & \oc{\gmZB} & \oc{\gmDC} & \oc{\gmDC} & \oc{{\bf 0}} \ps\\ \hld

 \oc{2B} & \oc{4A} & \oc{4B} & \oc{\gmZB} & \oc{\gmVA} & \oc{\gmVB} & \oc{$5 + \sqrt{5}$} \ps\\* \hl
 \oc{2B} & \oc{4A} & \oc{4F} & \oc{\gmZB} & \oc{\gmVA} & \oc{\gmVF} & \oc{$5 - \sqrt{5}$} \ps\\* \hl
 \oc{2B} & \oc{4B} & \oc{4B} & \oc{\gmZB} & \oc{\gmVB} & \oc{\gmVB} & \oc{$\frac{2}{79} \, (115 + 14 \, \sqrt{5})$} \ps\\* \hl
 \oc{2B} & \oc{4B} & \oc{4E} & \oc{\gmZB} & \oc{\gmVB} & \oc{\gmVE} & \oc{$5 + \sqrt{5}$} \ps\\* \hl
 \oc{2B} & \oc{4B} & \oc{4F} & \oc{\gmZB} & \oc{\gmVB} & \oc{\gmVF} & \oc{0} \ps\\* \hl
 \oc{2B} & \oc{4B} & \oc{4G} & \oc{\gmZB} & \oc{\gmVB} & \oc{\gmVG} & \oc{$5 + \sqrt{5}$} \ps\\* \hl
 \oc{2B} & \oc{4E} & \oc{4F} & \oc{\gmZB} & \oc{\gmVE} & \oc{\gmVF} & \oc{$5 - \sqrt{5}$} \ps\\* \hl
 \oc{2B} & \oc{4F} & \oc{4F} & \oc{\gmZB} & \oc{\gmVF} & \oc{\gmVF} & \oc{$\frac{2}{79} \, (115 - 14 \, \sqrt{5})$} \ps\\* \hl
 \oc{2B} & \oc{4F} & \oc{4G} & \oc{\gmZB} & \oc{\gmVF} & \oc{\gmVG} & \oc{$5 - \sqrt{5}$} \ps\\* \hl
 \oc{2B} & \oc{4G} & \oc{4G} & \oc{\gmZB} & \oc{\gmVG} & \oc{\gmVG} & \oc{0} \ps\\* \hld
 
 \oc{3B} & \oc{3B} & \oc{4A} & \oc{\gmDB} & \oc{\gmDB} & \oc{\gmVA} & \oc{${\bf \frac{1}{50} \, (261 + 9\, \sqrt{41})}$} \ps\\* \hl
 \oc{3B} & \oc{3B} & \oc{4B} & \oc{\gmDB} & \oc{\gmDB} & \oc{\gmVB} & \oc{$\frac{1}{11} \, (87 + 3\, \sqrt{5})$} \ps\\* \hl
 \oc{3B} & \oc{3B} & \oc{4E} & \oc{\gmDB} & \oc{\gmDB} & \oc{\gmVE} & \oc{${\bf \frac{1}{50} \, (261 - 9\, \sqrt{41})}$} \ps\\* \hl
 \oc{3B} & \oc{3B} & \oc{4F} & \oc{\gmDB} & \oc{\gmDB} & \oc{\gmVF} & \oc{$\frac{1}{11} \, (87 - 3\, \sqrt{5})$} \ps\\* \hl
 \oc{3B} & \oc{3C} & \oc{4B} & \oc{\gmDB} & \oc{\gmDC} & \oc{\gmVB} & \oc{$\frac{1}{11} \, (39 + 7\, \sqrt{5})$} \ps\\* \hl
 \oc{3B} & \oc{3C} & \oc{4F} & \oc{\gmDB} & \oc{\gmDC} & \oc{\gmVF} & \oc{$\frac{1}{11} \, (39 - 7\, \sqrt{5})$} \ps\\* \hl
 \oc{3B} & \oc{3C} & \oc{4G} & \oc{\gmDB} & \oc{\gmDC} & \oc{\gmVG} & \oc{4} \ps\\* \hl
 \oc{3C} & \oc{3C} & \oc{4A} & \oc{\gmDC} & \oc{\gmDC} & \oc{\gmVA} & \oc{${\bf \frac{1}{2} \, (13+\sqrt{41})}$} \ps\\* \hl
 \oc{3C} & \oc{3C} & \oc{4B} & \oc{\gmDC} & \oc{\gmDC} & \oc{\gmVB} & \oc{$5 + \sqrt{5}$} \ps\\* \hl
 \oc{3C} & \oc{3C} & \oc{4E} & \oc{\gmDC} & \oc{\gmDC} & \oc{\gmVE} & \oc{${\bf \frac{1}{2} \, (13-\sqrt{41})}$} \ps\\* \hl
 \oc{3C} & \oc{3C} & \oc{4F} & \oc{\gmDC} & \oc{\gmDC} & \oc{\gmVF} & \oc{$5 - \sqrt{5}$} \ps\\* \hl 
 \oc{3C} & \oc{3C} & \oc{4G} & \oc{\gmDC} & \oc{\gmDC} & \oc{\gmVG} & \oc{0} \ps\\* \hld
 
 \oc{4A} & \oc{4A} & \oc{4A} & \oc{\gmVA} & \oc{\gmVA} & \oc{\gmVA} & \oc{${\bf \frac{1}{733} \, (7185 + 309 \,\sqrt{41})}$} \ps\\* \hl
 \oc{4A} & \oc{4A} & \oc{4E} & \oc{\gmVA} & \oc{\gmVA} & \oc{\gmVE} & \oc{${\bf \frac{1}{10} \, (21 - \sqrt{41})}$} \ps\\* \hl
 \oc{4A} & \oc{4A} & \oc{4G} & \oc{\gmVA} & \oc{\gmVA} & \oc{\gmVG} & \oc{$\frac{1}{2} \, (13 + \sqrt{41})$} \ps\\* \hl
 \oc{4A} & \oc{4B} & \oc{4B} & \oc{\gmVA} & \oc{\gmVB} & \oc{\gmVB} & \oc{$12.3279656$} \ps\\* \hl
 \oc{4A} & \oc{4B} & \oc{4F} & \oc{\gmVA} & \oc{\gmVB} & \oc{\gmVF} & \oc{$\frac{1}{2} \, (9 + \sqrt{41})$} \ps\\* \hl
 \oc{4A} & \oc{4E} & \oc{4E} & \oc{\gmVA} & \oc{\gmVE} & \oc{\gmVE} & \oc{${\bf \frac{1}{10} \, (21 + \sqrt{41})}$} \ps\\* \hl
 \oc{4A} & \oc{4F} & \oc{4F} & \oc{\gmVA} & \oc{\gmVF} & \oc{\gmVF} & \oc{$4.865786$} \ps\\* \hl
 \oc{4A} & \oc{4G} & \oc{4G} & \oc{\gmVA} & \oc{\gmVG} & \oc{\gmVG} & \oc{$\frac{1}{2} \, (13 + \sqrt{41})$} \ps\\* \hl
 \oc{4B} & \oc{4B} & \oc{4B} & \oc{\gmVB} & \oc{\gmVB} & \oc{\gmVB} & \oc{$6.772955$} \ps\\* \hl
 \oc{4B} & \oc{4B} & \oc{4E} & \oc{\gmVB} & \oc{\gmVB} & \oc{\gmVE} & \oc{$38.020253$} \ps\\* \hl
  \oc{4B} & \oc{4B} & \oc{4F} & \oc{\gmVB} & \oc{\gmVB} & \oc{\gmVF} & \oc{$26.076638$} \ps\\* \hl
 \oc{4B} & \oc{4B} & \oc{4G} & \oc{\gmVB} & \oc{\gmVB} & \oc{\gmVG} & \oc{$\frac{4}{19} \, (25 + 7\, \sqrt{5})$} \ps\\* \hl
 \oc{4B} & \oc{4E} & \oc{4F} & \oc{\gmVB} & \oc{\gmVE} & \oc{\gmVF} & \oc{$\frac{1}{2} \, (9 - \sqrt{41})$} \ps\\* \hl
 \oc{4B} & \oc{4F} & \oc{4F} & \oc{\gmVB} & \oc{\gmVF} & \oc{\gmVF} & \oc{$5.374976$} \ps\\* \hl
 \oc{4B} & \oc{4F} & \oc{4G} & \oc{\gmVB} & \oc{\gmVF} & \oc{\gmVG} & \oc{$\frac{10}{3}$} \ps\\* \hl
 \oc{4B} & \oc{4G} & \oc{4G} & \oc{\gmVB} & \oc{\gmVG} & \oc{\gmVG} & \oc{$5 + \sqrt{5}$} \ps\\* \hl
 \oc{4E} & \oc{4E} & \oc{4E} & \oc{\gmVE} & \oc{\gmVE} & \oc{\gmVE} & \oc{${\bf \frac{1}{733} \, (7185  - 309 \,\sqrt{41})}$} \ps\\* \hl
 \oc{4E} & \oc{4E} & \oc{4G} & \oc{\gmVE} & \oc{\gmVE} & \oc{\gmVG} & \oc{$\frac{1}{2} \, (13 - \sqrt{41})$} \ps\\* \hl
 \oc{4E} & \oc{4F} & \oc{4F} & \oc{\gmVE} & \oc{\gmVF} & \oc{\gmVF} & \oc{$4.785995$} \ps\\* \hl
 \oc{4E} & \oc{4G} & \oc{4G} & \oc{\gmVE} & \oc{\gmVG} & \oc{\gmVG} & \oc{$\frac{1}{2} \, (13 - \sqrt{41})$} \ps\\* \hl
 \oc{4F} & \oc{4F} & \oc{4F} & \oc{\gmVF} & \oc{\gmVF} & \oc{\gmVF} & \oc{$4.464987$} \ps\\* \hl
 \oc{4F} & \oc{4F} & \oc{4G} & \oc{\gmVF} & \oc{\gmVF} & \oc{\gmVG} & \oc{$\frac{4}{19} \, (25 - 7\, \sqrt{5})$} \ps\\* \hl
 \oc{4F} & \oc{4G} & \oc{4G} & \oc{\gmVF} & \oc{\gmVG} & \oc{\gmVG} & \oc{$5 - \sqrt{5}$} \ps\\* \hl
 \oc{4G} & \oc{4G} & \oc{4G} & \oc{\gmVG} & \oc{\gmVG} & \oc{\gmVG} & \oc{0} \ps\\* \hld
 
 \oc{2B} & \oc{5J} & \oc{5J} & \oc{\gmZB} & \oc{\gmFJ} & \oc{\gmFJ} & \oc{$\frac{10}{7}$} \ps\\* \hl
 \oc{2B} & \oc{5J} & \oc{5K} & \oc{\gmZB} & \oc{\gmFJ} & \oc{\gmFK} & \oc{2} \ps\\* \hl
 \oc{2B} & \oc{5K} & \oc{5K} & \oc{\gmZB} & \oc{\gmFK} & \oc{\gmFK} & \oc{0} \ps\\* \hld
 
 \oc{3B} & \oc{4A} & \oc{5J} & \oc{\gmDB} & \oc{\gmVA} & \oc{\gmFJ} & \oc{2} \ps\\* \hl
 \oc{3B} & \oc{4A} & \oc{5K} & \oc{\gmDB} & \oc{\gmVA} & \oc{\gmFK} & \oc{4} \ps\\* \hl
 \oc{3B} & \oc{4B} & \oc{5J} & \oc{\gmDB} & \oc{\gmVB} & \oc{\gmFJ} & \oc{$\frac{2}{3} \, (13 + 2\, \sqrt{5})$} \ps\\* \hl
 \oc{3B} & \oc{4B} & \oc{5K} & \oc{\gmDB} & \oc{\gmVB} & \oc{\gmFK} & \oc{4} \ps\\* \hl
 \oc{3B} & \oc{4E} & \oc{5J} & \oc{\gmDB} & \oc{\gmVE} & \oc{\gmFJ} & \oc{2} \ps\\* \hl
 \oc{3B} & \oc{4E} & \oc{5K} & \oc{\gmDB} & \oc{\gmVE} & \oc{\gmFK} & \oc{4} \ps\\* \hl
 \oc{3B} & \oc{4F} & \oc{5J} & \oc{\gmDB} & \oc{\gmVF} & \oc{\gmFJ} & \oc{$\frac{2}{3} \, (13 - 2\, \sqrt{5})$} \ps\\* \hl
 \oc{3B} & \oc{4F} & \oc{5K} & \oc{\gmDB} & \oc{\gmVF} & \oc{\gmFK} & \oc{4} \ps\\* \hl
 \oc{3B} & \oc{4G} & \oc{5J} & \oc{\gmDB} & \oc{\gmVG} & \oc{\gmFJ} & \oc{2} \ps\\* \hl
 \oc{3B} & \oc{4G} & \oc{5K} & \oc{\gmDB} & \oc{\gmVG} & \oc{\gmFK} & \oc{4} \ps\\* \hl
 \oc{3C} & \oc{4A} & \oc{5J} & \oc{\gmDC} & \oc{\gmVA} & \oc{\gmFJ} & \oc{2} \ps\\* \hl
 \oc{3C} & \oc{4B} & \oc{5J} & \oc{\gmDC} & \oc{\gmVB} & \oc{\gmFJ} & \oc{2} \ps\\* \hl
 \oc{3C} & \oc{4B} & \oc{5K} & \oc{\gmDC} & \oc{\gmVB} & \oc{\gmFK} & \oc{$5+\sqrt{5}$} \ps\\* \hl
 \oc{3C} & \oc{4E} & \oc{5J} & \oc{\gmDC} & \oc{\gmVE} & \oc{\gmFJ} & \oc{2} \ps\\* \hl
 \oc{3C} & \oc{4F} & \oc{5J} & \oc{\gmDC} & \oc{\gmVF} & \oc{\gmFJ} & \oc{2} \ps\\* \hl
 \oc{3C} & \oc{4F} & \oc{5K} & \oc{\gmDC} & \oc{\gmVF} & \oc{\gmFK} & \oc{$5-\sqrt{5}$} \ps\\* \hl
 \oc{3C} & \oc{4G} & \oc{5J} & \oc{\gmDC} & \oc{\gmVG} & \oc{\gmFJ} & \oc{2} \ps\\* \hl
 \oc{3C} & \oc{4G} & \oc{5K} & \oc{\gmDC} & \oc{\gmVG} & \oc{\gmFK} & \oc{0} \ps\\* \hl
 \end{longtable}\end{center}

As reported in the introduction we make the general observation, that for a three-point
function of two protected operators with one unprotected operator the structure constants
follow the simple pattern:
\be
\frac{C^{(1)}_{\alpha \beta \gamma}}{C^{(0)}_{\alpha \beta \gamma}}=
-\frac{1}{2}\, \gamma_{\gamma}\, \qquad \text{if } \gamma_{\alpha}=\gamma_{\beta}=0\, .
\ee
This occurred in all applicable 17 cases in the above.
We again stress that, except for the cases of $\mathcal{K}$ and 3B, this result will 
generically receive corrections from subleading
operator-mixing terms with two fermion and two derivative insertions.

\subsection{Konishi operator with two primary scalar operators of
arbitrary lengths}

We calculated the three-point function of a Konishi operator with two arbitrary operators of same length from a diagonal basis. The three-point function then takes the general form
\begin{align}
 C^{(1)}_{\alpha \beta \K} &= - \kla{\frac{\gamma_\alpha}{\Dn_\alpha} + \frac{\gamma_\beta}{\Dn_\beta} + \frac{\gamma_\K}{\Dn_\K}} C^{(0)}_{\alpha \beta \K} = -\frac{\delta_{\alpha \beta}}{4 \pi^2 \, \sqrt{3}} \kla{2 \gamma_\alpha + \frac{3}{8 \pi^2} \Dn_\alpha}\, ,
\end{align}
as already mentioned in the introduction.

This may be shown as follows.
Let $\K$ be the length two Konishi operator and the set $\{\Op_\alpha\}$ an arbitrary non-diagonal basis for the operators of length $\Delta^{(0)}$ that can be written in terms of attached vectors, namely
\begin{align}
 \K &= \frac{1}{\sqrt{12}} \sum_i \tr{\phi^i \phi^i} \\
 \Op_\alpha &= \tr{u^\alpha_1 \cdot \phi \cdots u^\alpha_{\oplength} \cdot \phi} & (\oplength > 2).
\end{align}
Let $Z_k \subset S_k$ denote the set of cyclic permutations of $(1,2,\dots,k)$.

We choose the renormalization scheme $\e \to e \e$ in which only the 2-gons hold finite contributions
\begin{equation}
 \skla{\begin{minipage}[h]{12mm}\begin{center}\footnotesize

\begin{fmffile}{2gon}
\begin{fmfgraph*}(5,14)
\fmfstraight
\fmfleft{dummy1,p1,a2,a1,p2,dummy2}
\fmfright{dummy3,p3,b1,b2,p4,dummy4}
\fmfpen{thin}
\fmf{plain}{dummy1,p1,a2,a1,p2,dummy2}
\fmf{plain}{dummy3,p3,b1,b2,p4,dummy4}
\fmffreeze
\fmf{plain,width=2}{a1,b2}
\fmf{plain,width=2}{a2,b1}
\fmf{dots}{p1,p3}
\fmf{dots}{p2,p4}
\fmfdot{a1,a2,b1,b2}
\fmfv{label=$u_1$,label.dist=1.5thick}{a1}
\fmfv{label=$u_2$,label.dist=1.5thick}{a2}
\fmfv{label=$v_1$,label.dist=1.5thick}{b1}
\fmfv{label=$v_2$,label.dist=1.5thick}{b2}
\fmflabel{$x_1$}{dummy1}
\fmflabel{$x_2$}{dummy3}
\end{fmfgraph*}
\end{fmffile}

\end{center}\end{minipage}}_\text{1-loop} = I_{12}^2 \, \frac{\lambda}{8 \pi^2} \kla{\ln \frac{\e^2}{x_{12}^2} + 1} \Bigg( \begin{minipage}[h]{8mm}\begin{center}\footnotesize

\begin{fmffile}{2gonBlank}
\begin{fmfgraph*}(5,14)
\fmfstraight
\fmfleft{dummy1,p1,a2,a1,p2,dummy2}
\fmfright{dummy3,p3,b1,b2,p4,dummy4}
\fmfpen{thin}
\fmf{plain}{dummy1,p1,a2,a1,p2,dummy2}
\fmf{plain}{dummy3,p3,b1,b2,p4,dummy4}
\fmffreeze
\fmf{plain,width=2}{a1,b2}
\fmf{plain,width=2}{a2,b1}
\fmf{dots}{p1,p3}
\fmf{dots}{p2,p4}
\fmfdot{a1,a2,b1,b2}
\end{fmfgraph*}
\end{fmffile}

\end{center}\end{minipage} - \begin{minipage}[h]{8mm}\begin{center}\footnotesize

\begin{fmffile}{2gonCrossed}
\begin{fmfgraph*}(5,14)
\fmfstraight
\fmfleft{dummy1,p1,a2,a1,p2,dummy2}
\fmfright{dummy3,p3,b1,b2,p4,dummy4}
\fmfpen{thin}
\fmf{plain}{dummy1,p1,a2,a1,p2,dummy2}
\fmf{plain}{dummy3,p3,b1,b2,p4,dummy4}
\fmffreeze
\fmf{plain,width=2}{a1,b1}
\fmf{plain,width=2}{a2,b2}
\fmf{dots}{p1,p3}
\fmf{dots}{p2,p4}
\fmfdot{a1,a2,b1,b2}
\end{fmfgraph*}
\end{fmffile}

\end{center}\end{minipage} + \frac{1}{2} \begin{minipage}[h]{8mm}\begin{center}\footnotesize

\begin{fmffile}{2gonSelf}
\begin{fmfgraph*}(5,14)
\fmfstraight
\fmfleft{dummy1,p1,a2,a1,p2,dummy2}
\fmfright{dummy3,p3,b1,b2,p4,dummy4}
\fmfpen{thin}
\fmf{plain}{dummy1,p1,a2,a1,p2,dummy2}
\fmf{plain}{dummy3,p3,b1,b2,p4,dummy4}
\fmfcurved
\fmf{plain,left,width=2,tension=0}{a1,a2}
\fmf{plain,left,width=2,tension=0}{b1,b2}
\fmf{dots}{p1,p3}
\fmf{dots}{p2,p4}
\fmfdot{a1,a2,b1,b2}
\end{fmfgraph*}
\end{fmffile}

\end{center}\end{minipage} \Bigg)\vspace{5pt}
\end{equation}
while the 3-gons only contribute to the logarithmic terms. For the two-point functions we get
\begin{align}
 \skla{\Op_\alpha(x_1) \, \Op_\beta(x_2)} &= I_{12}^{\oplength} \sum_{\sigma \in Z_\oplength} \Bigg[ \prod_{i=1}^{\oplength} u^\alpha_i \cdot u^\beta_{\sigma(i)} + \frac{\lambda}{8 \pi^2} \kla{\lnk{12}+1}\nnl
 &\bleq \times \sum_{\tau \in Z_\oplength} \Big( u^\alpha_{\tau(1)} \cdot u^\beta_{\tau \circ \sigma(1)} \, u^\alpha_{\tau(2)} \cdot u^\beta_{\tau \circ \sigma(2)} - u^\alpha_{\tau(1)} \cdot u^\beta_{\tau \circ \sigma(2)} \, \nnl 
 &\bleq \times u^\alpha_{\tau(2)} \cdot u^\beta_{\tau \circ \sigma(1)} + \frac{1}{2} \, u^\alpha_{\tau(1)} \cdot u^\alpha_{\tau(2)} \, u^\beta_{\tau \circ \sigma(1)} \cdot u^\beta_{\tau \circ \sigma(2)} \Big) \nnl
 &\bleq \times \prod_{i=3}^{\oplength} u^\alpha_{\tau(i)} \cdot u^\beta_{\tau \circ \sigma(i)} \Bigg].
\end{align}

\noindent Now let $\D_\alpha = M_{\alpha \beta} \, \Op_\beta$ denote a diagonal basis of the length $\oplength$ subspace. Then
\begin{align}
 \skla{\D_\alpha(x_1) \, \D_\beta(x_2)} &= \frac{1}{x_{12}^{2 \oplength}} \kla{\delta_{\alpha \beta} + \lambda g_{\alpha \beta} + \lambda \gamma_\alpha \delta_{\alpha \beta} \lnk{12}} 
 = M_{\alpha \gamma} \, M_{\beta \delta} \, \skla{\Op_\gamma(x_1) \, \Op_\delta(x_2)}
\end{align}
from which we immediately get the condition for tree-level diagonality
\begin{equation}
 \sum_{\sigma \in Z_\oplength} M_{\alpha \gamma} \, M_{\beta \delta} \,\prod_{i=1}^{\oplength} u^\gamma_i \cdot u^\delta_{\sigma(i)} = (2 \pi)^{2 \oplength} \, \delta_{\alpha \beta}. \label{eqn:konishi proof tree-level diagonality}
\end{equation}
Using this result we obtain
\begin{align}
 \skla{\D_\alpha(x_1) \, \D_\beta(x_2)} &= \frac{1}{x_{12}^{2 \oplength}} \Bigg( \delta_{\alpha \beta} + \frac{\lambda}{8 \pi^2} \kla{\lnk{12}+1} \Bigg[ \oplength \, \delta_{\alpha \beta} - \frac{1}{(2 \pi)^{2 \oplength}} \nnl 
 &\bleq \times \sum_{\sigma \in Z_\oplength} \sum_{\tau \in Z_\oplength} M_{\alpha \gamma} \, M_{\beta \delta} \Big( u^\gamma_{\tau(1)} \cdot u^\delta_{\tau \circ \sigma(2)} \, u^\gamma_{\tau(2)} \cdot u^\delta_{\tau \circ \sigma(1)} \nnl 
 &\bleq - \frac{1}{2} u^\gamma_{\tau(1)} \cdot u^\gamma_{\tau(2)} \, u^\delta_{\tau \circ \sigma(1)} \cdot u^\delta_{\tau \circ \sigma(2)} \Big) \times \prod_{i=3}^{\oplength} u^\gamma_{\tau(i)} \cdot u^\delta_{\tau \circ \sigma(i)} \Bigg] \Bigg)
\end{align}
and thus the condition for one-loop diagonality
\begin{align}
(2 \pi)^{2 \oplength} \, \delta_{\alpha \beta} \, \kla{\oplength - 8 \pi^2 \, \gamma_\alpha} &= \sum_{\sigma \in Z_\oplength} \sum_{\tau \in Z_\oplength} M_{\alpha \gamma} \, M_{\beta \delta} \Big( u^\gamma_{\tau(1)} \cdot u^\delta_{\tau \circ \sigma(2)} \, u^\gamma_{\tau(2)} \cdot u^\delta_{\tau \circ \sigma(1)} \nnl &
 \bleq - \frac{1}{2} u^\gamma_{\tau(1)} \cdot u^\gamma_{\tau(2)} \, u^\delta_{\tau \circ \sigma(1)} \cdot u^\delta_{\tau \circ \sigma(2)} \Big) \,  \prod_{i=3}^{\oplength} u^\gamma_{\tau(i)} \cdot u^\delta_{\tau \circ \sigma(i)} \label{eqn:konishi proof one-loop diagonality}
\intertext{and}
 g_\alpha &= \gamma_\alpha \, . \label{eqn:g alpha beta in eps to eeps renormalization}
\end{align}

\noindent The three-point functions are
\begin{align}
 &\hspace{-20pt}\skla{\D_\alpha(x_1) \, \D_\beta(x_2) \K(x_3)} = M_{\alpha \gamma} \, M_{\beta \delta} \, \skla{\Op_\alpha(x_1) \, \Op_\beta(x_2) \K(x_3)} \nnl
 &= \frac{1}{(2 \pi)^{2 \oplength +2} \, \sqrt{3} \, x_{12}^{2 \oplength -2} \, x_{13}^2 \, x_{23}^2} \, \sum_{\sigma \in Z_\oplength} \sum_{\tau \in Z_\oplength} M_{\alpha \gamma} \, M_{\beta \delta}  \times \Bigg[ \prod_{i=1}^{\oplength} u^\gamma_{\sigma(i)} \cdot u^\delta_{\tau(i)} \nnl
 &\bleq+ \frac{\lambda}{8 \pi^2} \sum_{\rho \in Z_{\oplength-2}} \Big( u^\gamma_{\sigma \circ \rho(1)} \cdot u^\delta_{\tau \circ \rho(1)} \, u^\gamma_{\sigma \circ \rho(2)} \cdot u^\delta_{\tau \circ \rho(2)} - u^\gamma_{\sigma \circ \rho(1)} \cdot u^\delta_{\tau \circ \rho(2)} \, u^\gamma_{\sigma \circ \rho(2)} \cdot u^\delta_{\tau \circ \rho(1)} \nnl
 &\bleq+ \frac{1}{2} u^\gamma_{\sigma \circ \rho(1)} \cdot u^\gamma_{\sigma \circ \rho(2)} \, u^\delta_{\tau \circ \rho(1)} \cdot u^\delta_{\tau \circ \rho(2)} \Big) \times \prod_{i=3}^{\oplength -2} \kla{u^\gamma_{\sigma \circ \rho(i)} \cdot u^\delta_{\tau \circ \rho(i)}} \nnl
 &\bleq\times u^\gamma_{\sigma(\oplength-1)} \cdot u^\delta_{\tau(\oplength-1)} \, u^\gamma_{\sigma(\oplength)} \cdot u^\delta_{\tau(\oplength)} + \lambda \times \text{logs} \Bigg]\nnl
 &\soll \frac{1}{x_{12}^{2 \oplength -2} \, x_{13}^2 \, x_{23}^2} \kla{C^{(0)}_{\alpha \beta \K} + \lambda \, \widetilde{C}^{(1)}_{\alpha \beta \K} + \lambda \times \text{logs}}
\end{align}
and we obtain the tree-level structure constant
{\allowdisplaybreaks \begin{align}
 C^{(0)}_{\alpha \beta \K} &= \frac{1}{(2 \pi)^{2 \oplength +2} \, \sqrt{3}} \sum_{\sigma \in Z_\oplength} \sum_{\tau \in Z_\oplength} M_{\alpha \gamma} \, M_{\beta \delta} \prod_{i=1}^{\oplength} u^\gamma_{\sigma(i)} \cdot u^\delta_{\tau(i)} \nnl
 &= \frac{\oplength}{(2 \pi)^{2 \oplength +2} \, \sqrt{3}} \sum_{\tau \in Z_\oplength} M_{\alpha \gamma} \, M_{\beta \delta} \prod_{i=1}^{\oplength} u^\gamma_{i} \cdot u^\delta_{\tau(i)},
\end{align}
}where we omitted one sum over all permutations in the second line because the first sum already delivers all possible contractions.

Using equation (\ref{eqn:konishi proof tree-level diagonality}) we get
\begin{equation}
C^{(0)}_{\alpha \beta \K} = \frac{\Dn}{4 \pi^2 \, \sqrt{3}} \, \delta_{\alpha \beta}\, . 
\label{a1}
\end{equation}

\noindent The one-loop structure constant is
\begin{small}\begin{align}
 \widetilde{C}^{(1)}_{\alpha \beta \K} &= \frac{1}{(2 \pi)^{2 \oplength +4} \, \sqrt{12}} \sum_{\sigma \in Z_\oplength} \sum_{\tau \in Z_\oplength} \sum_{\rho \in Z_{\oplength-2}} M_{\alpha \gamma} \, M_{\beta \delta} \nnl
 &\bleq \times \Bigg[ \prod_{i=1}^{\oplength-2} \kla{u^\gamma_{\sigma \circ \rho(i)} \cdot u^\delta_{\tau \circ \rho(i)}} \times u^\gamma_{\sigma(\oplength-1)} \cdot u^\delta_{\tau(\oplength-1)} \, u^\gamma_{\sigma(\oplength)} \cdot u^\delta_{\tau(\oplength)} \nnl
 &\bleq - \Big( u^\gamma_{\sigma \circ \rho(1)} \cdot u^\delta_{\tau \circ \rho(2)} \, u^\gamma_{\sigma \circ \rho(2)} \cdot u^\delta_{\tau \circ \rho(1)} - \frac{1}{2} \, u^\gamma_{\sigma \circ \rho(1)} \cdot u^\gamma_{\sigma \circ \rho(2)} u^\delta_{\tau \circ \rho(1)} \cdot u^\delta_{\tau \circ \rho(2)} \Big) \nnl
 &\bleq \times \prod_{i=3}^{\oplength-2} \kla{u^\gamma_{\sigma \circ \rho(i)} \cdot u^\delta_{\tau \circ \rho(i)}} \times u^\gamma_{\sigma(\oplength-1)} \cdot u^\delta_{\tau(\oplength-1)} \, u^\gamma_{\sigma(\oplength)} \cdot u^\delta_{\tau(\oplength)} \Bigg] \nnl
 &= \frac{\delta_{\alpha \beta}}{(2 \pi)^{4} \, \sqrt{12}} \ekla{ (\oplength-2) \, \oplength  - (\oplength-2) \, (\oplength - 8 \pi^2 \, \gamma_\alpha) } \nnl
 &= \frac{(\oplength-2) \, \gamma_\alpha}{4 \pi^2 \, \sqrt{3}} \, \delta_{\alpha \beta},
 \label{a2}
\end{align}
\end{small}where the sum over the $\rho$-permutations gives only a factor of $(\oplength-2)$ and we made use of equations (\ref{eqn:konishi proof tree-level diagonality}) and (\ref{eqn:konishi proof one-loop diagonality}) in the second step.

The renormalization scheme independent structure constants 
\be
 C^{(1)}_{\alpha \beta \gamma} = \widetilde{C}^{(1)}_{\alpha \beta \gamma} - \frac{1}{2} \, C^{(0)}_{\alpha \beta \gamma} \, \kla{g_\alpha  + g_\beta + g_\gamma}
\ee
may now be written down using \eqn{a1}, \eqn{a2} and 
(\ref{eqn:g alpha beta in eps to eeps renormalization})
to find
\be
 C^{(1)}_{\alpha \beta \K} = \widetilde{C}^{(1)}_{\alpha \beta \K} - \frac{1}{2} \, C^{(0)}_{\alpha \beta \K} \, \kla{\gamma_\alpha + \gamma_\beta + \frac{3}{4 \pi^2}} 
 =  - \kla{ \frac{\gamma_\alpha}{\oplength_\alpha} + \frac{\gamma_\beta}{\oplength_\beta} + \frac{\gamma_\K}{\oplength_\K} } \, C^{(0)}_{\alpha \beta \K} \, . \label{eqn:konishi beweis ergebnis}
\ee

\subsection*{Acknowledgements}

We thank Gleb Arutyunov, Niklas Beisert, Harald Dorn, George Georgiou, Valeria Gili,
Johannes Henn, Charlotte Kristjansen and Rodolfo Russo  
for helpful discussions. This work was supported by the Volkswagen Foundation.



\bibliographystyle{nb}
\bibliography{botany}

\end{document}